\documentclass[prd,showpacs,amssymb,amsmath,twocolumn]{revtex4}
\usepackage{bm}
\usepackage{graphicx}

\newcommand\rhat{\hat{\bf r}}
\newcommand\khat{\hat{\bf k}}

\newcommand\nhat{\hat{\bf n}}

\newcommand\etarec{\eta_{\rm rec}}

\newcommand\Ytwo{{}_2Y}
\newcommand\beq{\begin{equation}}
\newcommand\eeq{\end{equation}}

\begin{document}
\title{Probing the Universe on Gigaparsec Scales with Remote \\
Cosmic Microwave
Background Quadrupole Measurements}
\author{Emory F. Bunn}
\email{ebunn@richmond.edu}
\affiliation{Physics Department, University of Richmond, Richmond, VA 23173}
\date{May 24, 2006}
\begin{abstract}
Scattering of cosmic microwave background (CMB) radiation in galaxy
clusters induces a polarization signal proportional to the CMB
quadrupole anisotropy at the cluster's location and look-back time.  A survey
of such remote quadrupole measurements provides
information about large-scale cosmological perturbations.
This paper presents a formalism for calculating the correlation
function of remote quadrupole measurements in spherical harmonic
space. The number of independent modes probed by both single-redshift
and volume-limited surveys 
is presented, along with the length scales probed by these modes.
In a remote quadrupole survey sparsely covering a large area of
sky, the largest-scale modes probe the same length scales
as the quadrupole but with much narrower Fourier-space window functions.
The largest-scale modes are significantly correlated with the local
CMB, but even when this correlation is projected out the
largest remaining modes probe gigaparsec scales (comparable
to the CMB at $l=2$-10) with narrow window functions.
These modes may provide insight
into the possible anomalies in the large-scale CMB anisotropy.
At fixed redshift, the data from such a survey 
form an $E$-type spin-2 field on the
sphere to a good approximation; the near-absence 
of $B$ modes will provide a valuable
check on systematic errors.
A survey of only a few low-redshift clusters allows an independent
reconstruction of the five coefficients of the local CMB quadrupole,
providing a test for contamination in the WMAP quadrupole.
The formalism presented here is also useful for analyzing
smaller-scale surveys to probe the late integrated Sachs-Wolfe effect
and hence the properties of dark energy.
\end{abstract}


\maketitle

\section{Introduction}

Cosmologists are making rapid progress in our understanding
of the structure of the large-scale Universe.  On the largest
scales, the chief source of information is the cosmic microwave background
(CMB) anisotropy and polarization, 
particularly the all-sky data from WMAP 
\cite{wmapbasic,wmapspectrum,wmappol,wmap3basic,wmap3pol,wmap3imp}.
There have been tantalizing hints of unexpected behavior in the
largest-scale modes of the CMB anisotropy.  The COBE DMR
detected a lack of anisotropy power on the largest angular scales
\cite{bennettcobe,gorskicobe,hinshawcobe}, 
and WMAP has confirmed this result \cite{wmapbasic,wmapspectrum,wmap3basic}.
There is evidence suggesting that these largest-scale modes
are inconsistent with statistically isotropic theories
because of correlations between modes and/or asymmetry between
hemispheres
\cite{costaetal,copi,schwarz,hansen,land,bernui2,bielewicz,copi2,bernui};
however, there is disagreement over how to interpret these
results \cite{land,bielewicz,efstathiou}.  
In particular, these results are subject
to the classic problem of a posteriori interpretation of statistical
significances: if an unexpected anomaly is found, and its statistical
significance is computed thereafter, one cannot necessarily 
take the significance
at face value.  (After all, in any large data set, something unlikely
is bound to occur.)

The best way to resolve this situation is of course to obtain a new,
independent data set probing the same physical scales.  Unfortunately,
large-angular-scale CMB
observations are already at the ``cosmic variance'' limit,
and other 
independent probes of these ultra-large scales are few.  Observations
that may provide independent information on these scales
are therefore of considerable interest.
Large-angular-scale CMB polarization data provide some relevant
information \cite{dore,skordis}, although the number of independent modes
probed is small and the results may depend on the details of reionization.

The scattering of CMB photons in clusters of galaxies may shed light
on this puzzle.
This scattering induces a polarization signal \cite{sazonov}, which is
determined by the quadrupole anisotropy in the photon distribution at
the cluster location.  This ``remote quadrupole'' 
signal probes large-scale modes of the
density perturbation field that are different from those probed by the
local CMB, so by measuring these remote quadrupoles it may be possible to
get around the cosmic variance limit \cite{kamionloeb}.
It has therefore been proposed
that a survey of remote quadrupoles may shed light on
the puzzle of large-scale CMB anomalies \cite{seto,baumanncooray}.

The remote quadrupole signals from different clusters are strongly
correlated with
each other and with the local CMB anisotropy \cite{portsmouth}.  
It is therefore not obvious how to design a survey to obtain
the maximum amount of new information.  In addition, we wish
to know what physical scales of perturbation are probed
by a given survey; this will depend on both the redshifts
and the angular distribution of clusters observed.
In this article I will develop a formalism for determining the independent
fluctuation modes that are probed by a survey of remote quadrupoles.
For a survey that sparsely covers a large area of sky,
the largest-scale modes probe comparable length scales to the first
few CMB
multipoles, with Fourier-space window functions that are narrower
than that of the local CMB.  Determination of these modes
may be expected to provide insight into the interpretation
of the possible anomalies in the large-scale CMB observations.
The value of a sparse large-area survey for this purpose
has been noted elsewhere \cite{seto}.  This
paper provides the first detailed assessment of the amount
of information available in such a survey.

The correlation function of remote quadrupole measurements
is quite complicated, depending on both the clusters' redshifts
and their angular separation \cite{portsmouth}.  At fixed redshift,
the remote quadrupole is a spin-2 field on the sky, so it
is natural to express it as an expansion in spin-2 spherical
harmonics.  Because it is predominantly derived from scalar perturbations,
at any given redshift it contains (to a good
approximation) only $E$ modes, with no $B$ contribution.
This should provide a valuable check on systematic errors in any
future survey.

At low redshift, the measurements naturally
become strongly correlated with the local CMB temperature quadrupole.  As
a result, the five coefficients $a_{2m}$ of the local
quadrupole can be easily measured from a survey of only a few low-redshift
clusters \cite{baumanncooray}.

The spherical harmonic basis diagonalizes the angular
correlations, giving a sequence of correlation functions that depend
only on redshift.  
It is much simpler to determine and count
the independent normal modes in the spherical harmonic 
basis rather than in real space.
For a survey that covers only part of the sky, of course, the
individual spherical harmonic coefficients will not be measured.
However, just as in the case of the local CMB we can still use the
spherical harmonic basis to count the number
of modes that can be measured, scaling the results by the
fraction of sky covered.

On smaller scales, a remote quadrupole survey provides insight
into the growth of structure in the recent past 
\cite{baumanncooray,CoorayBaumann,CHB,seto}.
The remote quadrupole signal, like the local CMB,
contains contributions both from the surface of last scattering
and from the integrated Sachs-Wolfe (ISW) effect 
resulting from time variations in the gravitational potential along
the line of sight \cite{sachswolfe}.  (See, e.g., \cite{hureview} for an overview of the
physics of CMB anisotropy.)
Since the ISW
contribution to the remote quadrupole measurements differs
from that of the local quadrupole, it is possible to extract information
about the recent growth of perturbations.  The formalism
developed in this paper provides
a method of quantifying
the amount of extra information that can be obtained from such
a survey.

If a cluster has a peculiar velocity, then there is a kinematically
induced polarization signal as well as the
signal considered here \cite{challinorfordlasenby,shimon}.  This kinematic contribution can
be removed through multifrequency observations \cite{CoorayBaumann},
and will be ignored in this paper.  In addition, we will not consider 
the polarization induced by scattering off of diffuse structure
\cite{liudasilva}; rather, we will envision a survey directed
at specific clusters of known redshift.

This paper is structured as follows.  Section \ref{sec:formalism}
develops the formalism for calculating the correlation function
of remote quadrupole measurements.  Section \ref{sec:surveymodes}
shows the information that can be obtained in hypothetical
remote-quadrupole surveys on a shell at a single redshift as well
as in volume-limited surveys,  and
section \ref{sec:discussion} contains
a discussion of the significance of these results.  Some more
than usually boring mathematical steps are contained in an appendix.

\section{Formalism}
\label{sec:formalism}

\subsection{Remote quadrupole in a single cluster}

We will assume a flat spatial
geometry and label any cluster's comoving position
with an ordinary 3-vector ${\bf r}$, with spherical coordinates
$(r,\theta_{\rhat},\phi_{\rhat})$ 
defined in some fixed earth-centered coordinate system.

For any particular cluster,
we will find it convenient to introduce
a second coordinate system,
denoted by a prime,
which will have its $z'$ axis is aligned with $\rhat$, the direction
from earth to the cluster.
To be specific,
let the
primed coordinate system be obtained from
the unprimed by rotating through an angle $\theta_{\rhat}$ about the $x$
axis and then by an angle $\phi_{\rhat}$ about the (original) $z$ axis, with
the third Euler angle set to zero.

Suppose that an observer in that cluster
at the cluster look-back time measures the CMB anisotropy,
conveniently recording the results
using the primed coordinate system:
\beq
\frac{\Delta T}{T}(\nhat')=\sum_{l,m} a_{lm}({\bf r})
Y_{lm}(\nhat').
\eeq
The quadrupole spherical harmonic coefficients are
\beq
a_{2m}({\bf r})=\int d^3k\,\Delta_2(k;r)\delta_\Phi({\bf k})
e^{i{\bf k}\cdot{\bf r}}Y_{2m}^*(\khat').
\eeq
Here $\delta_\Phi$ is the Fourier-space perturbation in the
gravitational potential.  On the
large scales of interest to us, the quadrupole transfer function contains
Sachs-Wolfe and ISW terms:
\begin{widetext}
\beq
\Delta_2(k;r)=-\frac{4\pi}{3}\left(j_2[k(\eta-\etarec)]+6\int_{\etarec}^\eta
d\eta' j_2[k(\eta-\eta')]\frac{\partial}{\partial\eta'}
\left(\frac{D(\eta')}{a(\eta')}\right)\right).
\eeq
In this expression $j_2$ is a spherical Bessel function, $a$
is the scale factor normalized to unity today, $\eta=\eta_0-r$ is conformal
time ($d\eta=dt/a(t)$), $\etarec$ is the conformal time of recombination,
$\eta_0$ is conformal time today,
and $D$ is the matter perturbation growth factor normalized to unity
at high redshift (e.g., \cite{padmanabhan}).  We assume that the dark
energy is spatially uniform, e.g., a cosmological constant.
In addition, we assume instantaneous recombination and ignore reionization.  
Most of the quadrupole signal seen by observers in the cluster
is due to photons that come from last scattering (just as most of the
local quadrupole signal is), so neglecting reionization is a good approximation
in this context.  On the other hand, we cannot ignore reionization
when considering the correlation between the remote quadrupole signal
and the local CMB {\it polarization} quadrupole, as discussed below.
We work in units where $c=1$.

The observed cluster polarization signal is proportional to the
$m=\pm 2$ spherical harmonic coefficients:
\beq
p_\pm({\bf r})\equiv (Q\pm iU)({\bf r})=Na_{2\pm 2}
({\bf r}).
\eeq
where 
\beq
N=\sqrt{\frac{3}{40\pi}}\tau,
\eeq
and
$\tau$ is the cluster optical depth.  We want to study
the behavior of $p_\pm$ as a function of cluster position ${\bf r}$.
Since $p_\pm$ are complex conjugates of each other, we
need only compute one of them.  Let's focus on $p_-({\bf r})$, which 
we will call simply $p({\bf r})$ from now on.

The  observed signal is
\beq
p({\bf r})=Na_{2-2}({\bf r})
=
N\int d^3k\, \Delta_2(k; r)\delta_\Phi({\bf k})
e^{i{\bf k}\cdot{\bf r}}Y_{22}(\khat').
\label{eq:b}
\eeq
Using equation (\ref{Y22equation}), we can write $Y_{22}(\khat')$
in the unprimed coordinate system:
\beq
p({\bf r})=
N\sqrt{\frac{4\pi}{5}}
\sum_{m=-2}^2(-1)^m \Ytwo_{2-m}(\rhat)
\int d^3k\,\Delta_2(k; r)\delta_\Phi({\bf k})
e^{i{\bf k}\cdot{\bf r}}Y_{2m}(\khat),
\eeq
where $\Ytwo_{2-m}$ is a spin-2 spherical harmonic.

For a fixed distance $r$, $p$ is a spin-2 function of direction
$\rhat$, so it is natural to 
expand in spin-2 spherical
harmonics:
\beq
p({\bf r})=\sum_{L,M}p_{LM}(r)\ \Ytwo_{LM}(\rhat),
\label{eq:ylmexpansion}
\eeq
with coefficients given by
\begin{align}
p_{LM}(r)&=\int d^2\rhat\,p({\bf r})\ \Ytwo_{LM}^*\\
&=N\sqrt{\frac{4\pi}{5}}\sum_{m} (-1)^m
\int d^3k\,\Delta_2(k; r)\delta_\Phi({\bf k})Y_{2m}(\khat)
\int d^2\rhat\  \Ytwo_{2-m}(\rhat)\ \Ytwo_{LM}^*(\rhat) e^{i
{\bf k}\cdot{\bf r}}.
\label{blmequation}
\end{align}
By expanding the exponential in spherical harmonics as shown
in the Appendix,
we can express the coefficients in the following form:
\beq
p_{LM}(r)=i^L\int d^3k \,\Delta_2(k;r)\delta_\Phi({\bf k})
F_L(kr)Y_{LM}^*(\khat),
\label{eq:plm}
\eeq
where
\beq
F_L(x)=\sqrt{20\pi}\sum_{\lambda=L-2,L,L+2}
(-1)^{(\lambda-L)/2}(2\lambda+1)
\begin{pmatrix}
2 & L & \lambda\\
2 & -2 & 0
\end{pmatrix}
\begin{pmatrix}
2 & L & \lambda \\ 0 & 0 & 0
\end{pmatrix}
j_\lambda(x).
\eeq

\begin{figure*}
\includegraphics[width=3in]{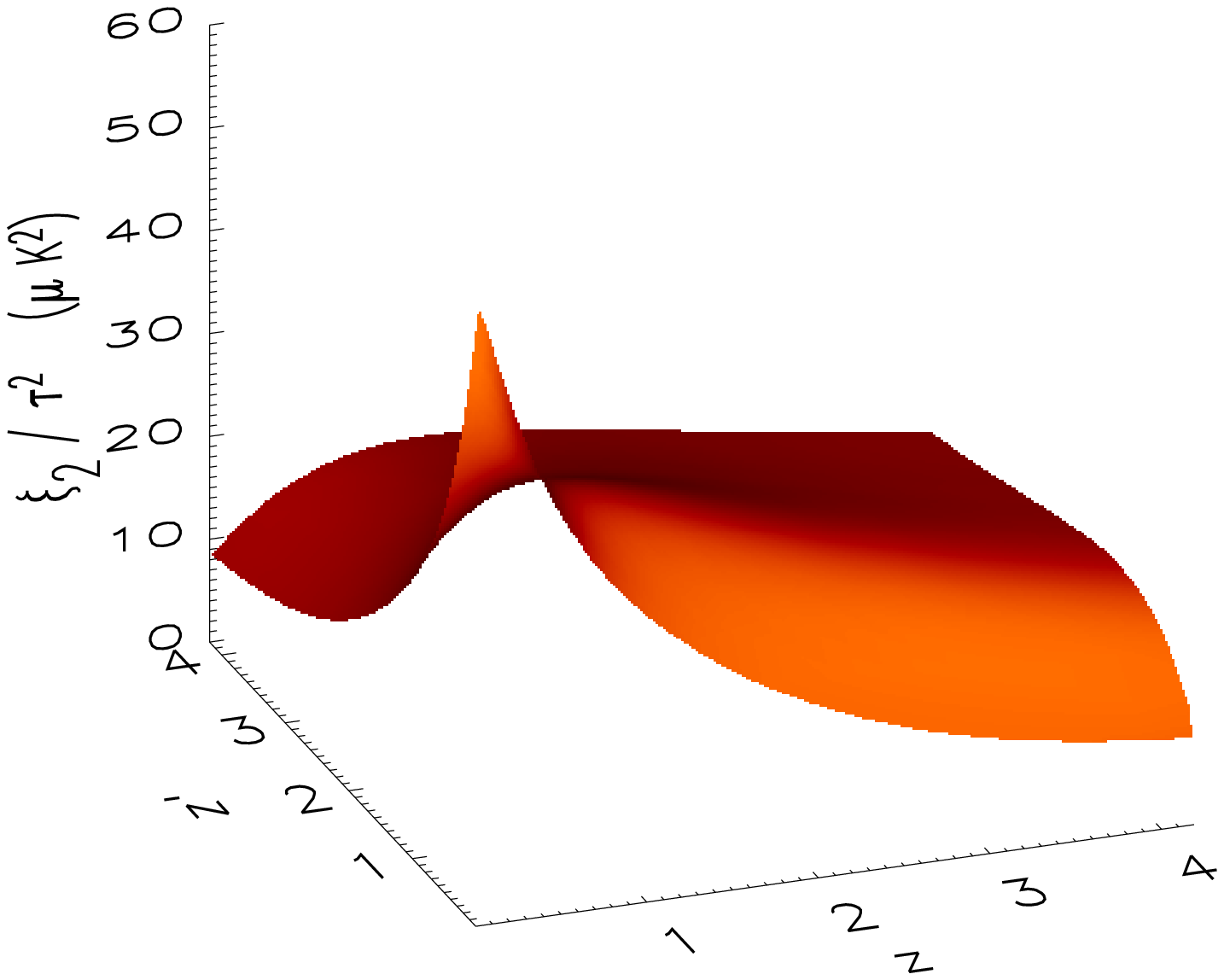}
\includegraphics[width=3in]{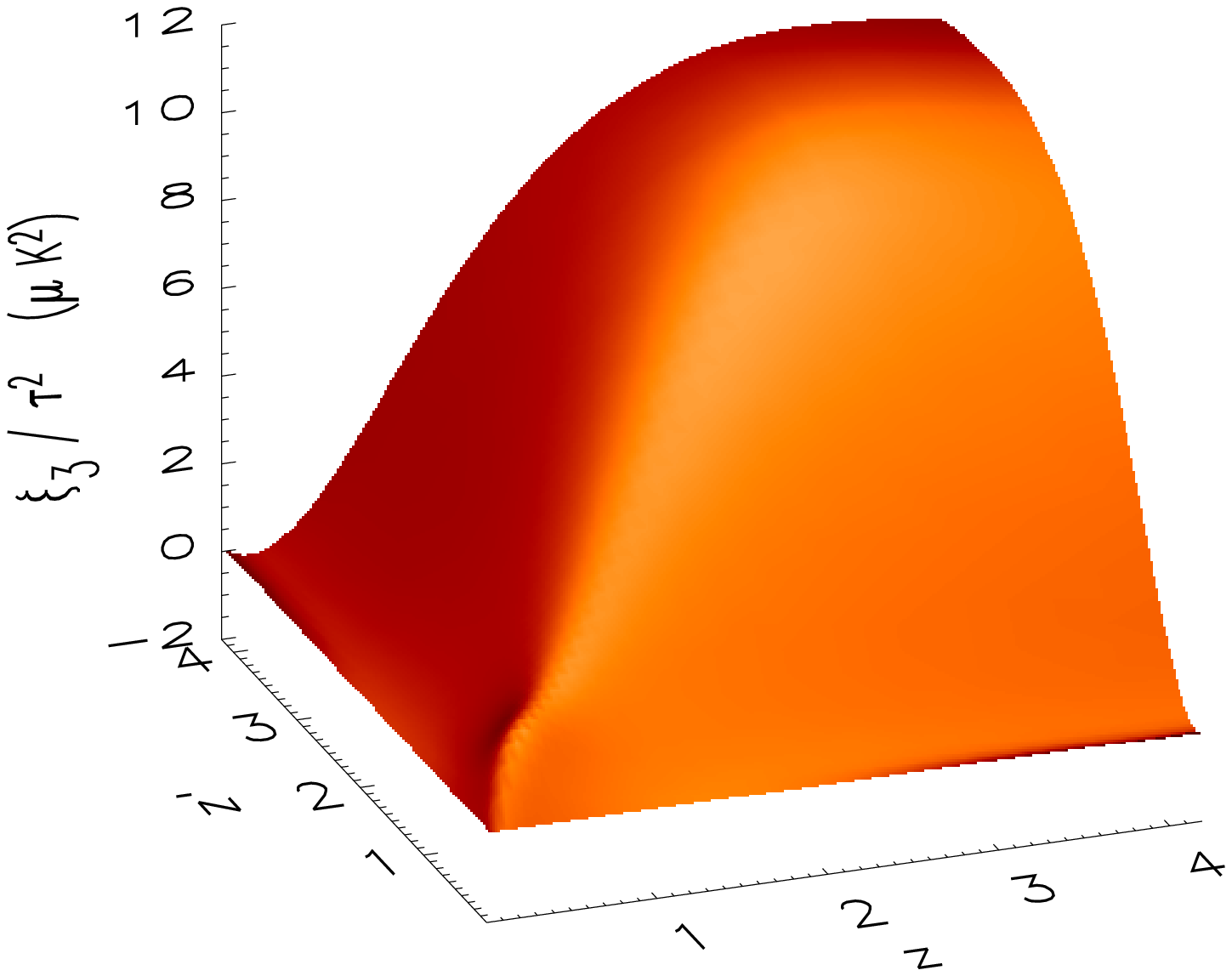}

\includegraphics[width=3in]{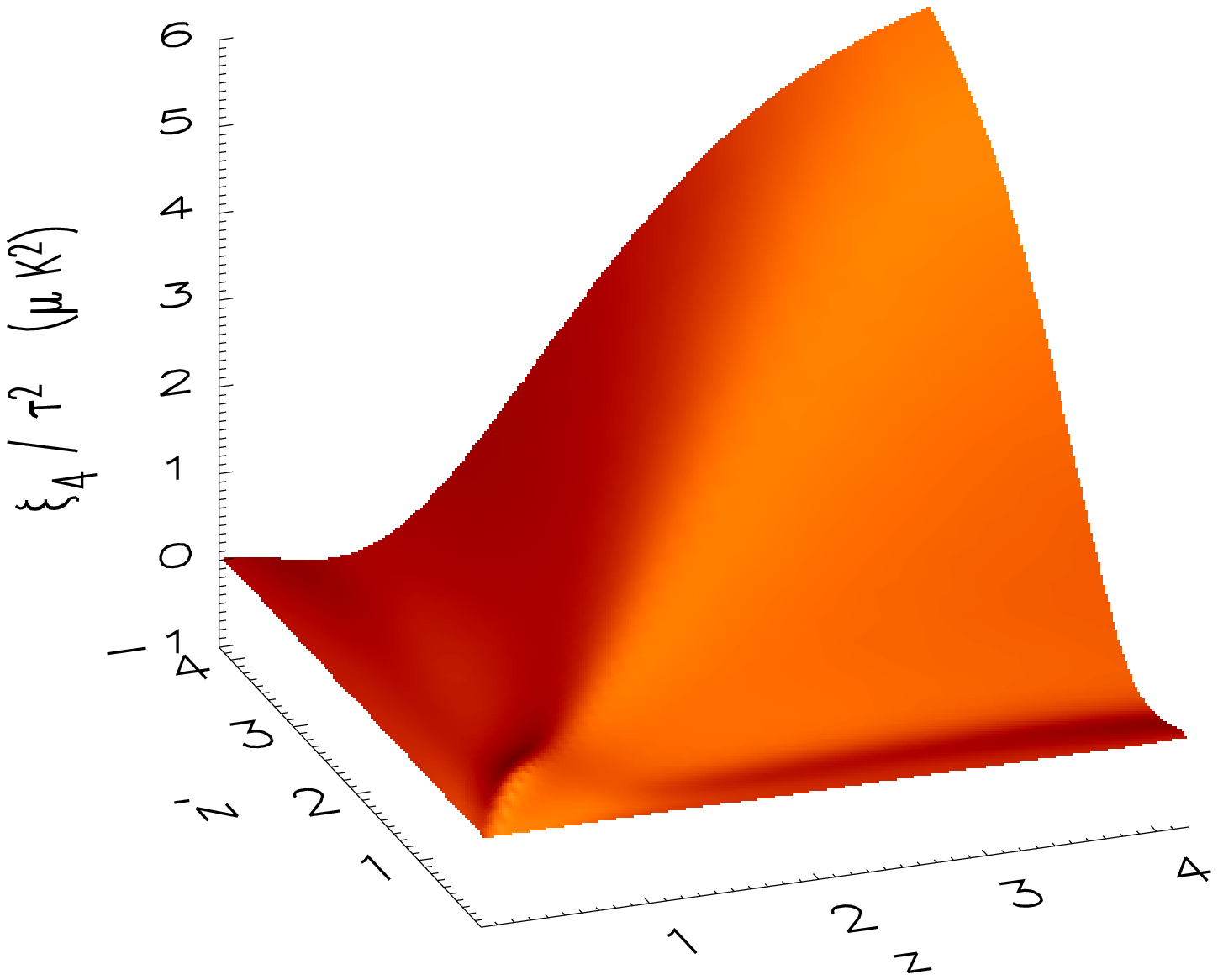}
\includegraphics[width=3in]{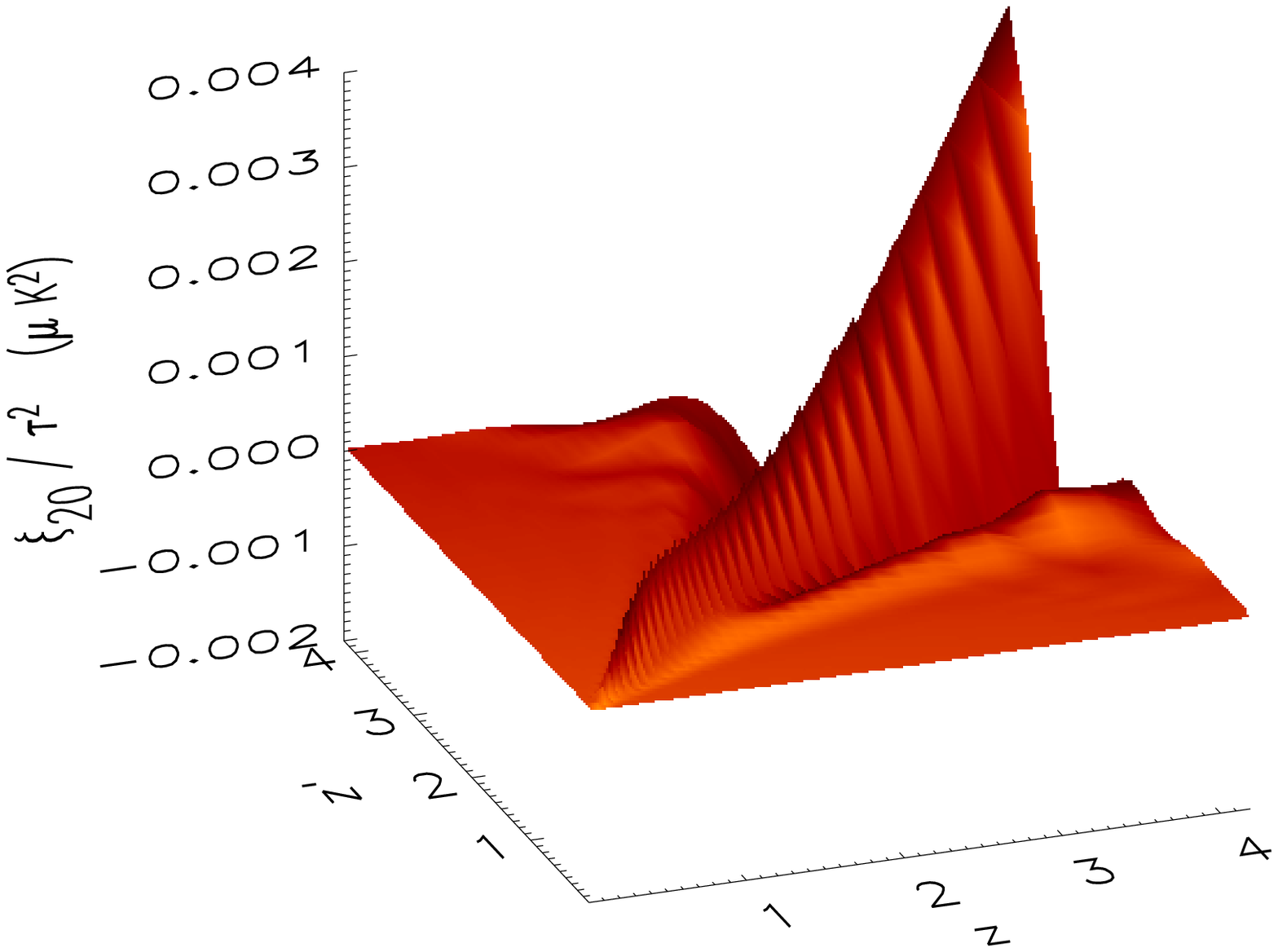}
\caption{Correlation functions $\xi_l$ for $l=2,3,4,20$.}
\label{fig:corrfuncs}
\end{figure*}

It is straightforward to check from equation (\ref{eq:plm})
that all
coefficients $p_{lm}(r)$ are real.  In the terminology of
CMB polarization \cite{eb1,eb2}, this means that the remote quadrupole
data form an $E$-type spin-2 field at any given distance,
with no $B$ modes.  The absence of $B$ modes arises because we have
considered only scalar perturbations as the source of the CMB quadrupole.
If tensor perturbations were included, then in principle a $B$
component would arise.  Considering the difficulty of 
detecting a remote quadrupole signal at all, the prospect of searching
for a subdominant $B$-type signal sounds extremely daunting.
It is probably more realistic, therefore, to search
for the scalar ($E$-type) signal in such a survey, using
the predicted absence
of $B$ modes as a check 
on systematic errors and noise (see Section \ref{sec:discussion}).

\end{widetext}
\subsection{Correlations between clusters}

Suppose that many clusters have been observed at many
different positions ${\bf r}_i$.
To determine the amount of information that can be obtained
from such a survey, we need to know the correlations $\langle
p({\bf r}_i)p({\bf r}_j)\rangle$.  The full correlation function
depends on
the directions as well as the distances of the clusters \cite{portsmouth}.
The correlation functions are simpler to work with in spherical
harmonic space, as the orthogonality of the spherical harmonics implies
that there are no correlations between $p_{LM}$ and $p_{L'M'}$:
\beq
\langle p_{LM}(r)p_{L'M'}^*(r')\rangle=\xi_{L}(r,r')\delta_{LL'}\delta_{MM'}.
\label{eq:plmcorr}
\eeq
The correlation function is
\begin{align}
&\xi_{L}(r,r')=\langle p_{LM}(r)p_{LM}^*(r')\rangle=\nonumber\\
&\int_0^\infty dk\,k^2\Delta_2(k;r)\Delta_2(k;r')F_L(kr)F_L(kr')P_\Phi(k),
\end{align}
where the power spectrum $P_\Phi$ is given by
\beq
\langle\delta_\Phi({\bf k})\delta^*_\Phi({\bf k}')\rangle=P_\Phi(k)
\delta^{(3)}({\bf k}-{\bf k}').
\eeq
On the scales of interest, $P_\Phi(k)\propto k^{n-4}$ where $n\approx
1$ is the spectral index.

\begin{figure*}
\includegraphics[width=3in]{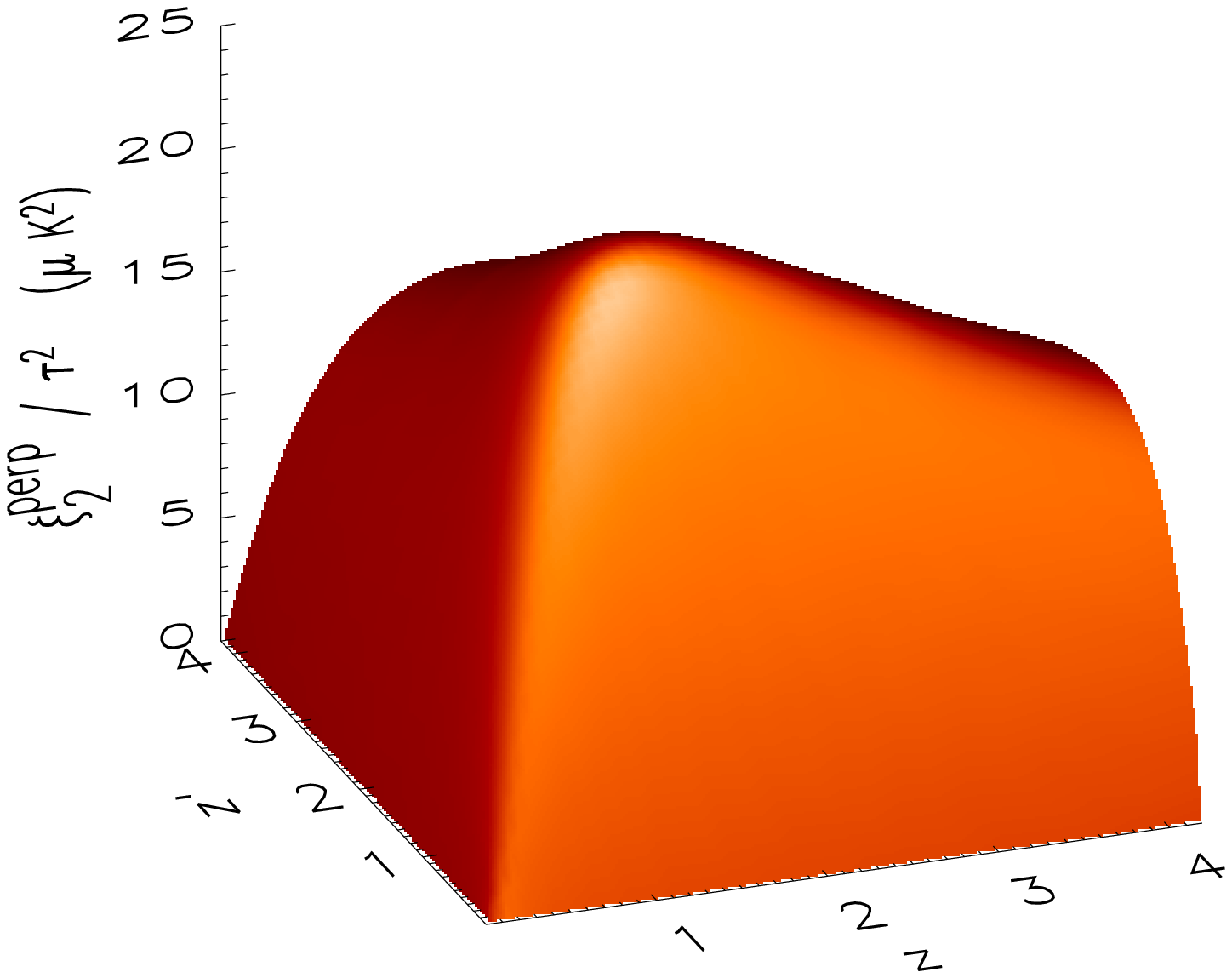}
\includegraphics[width=3in]{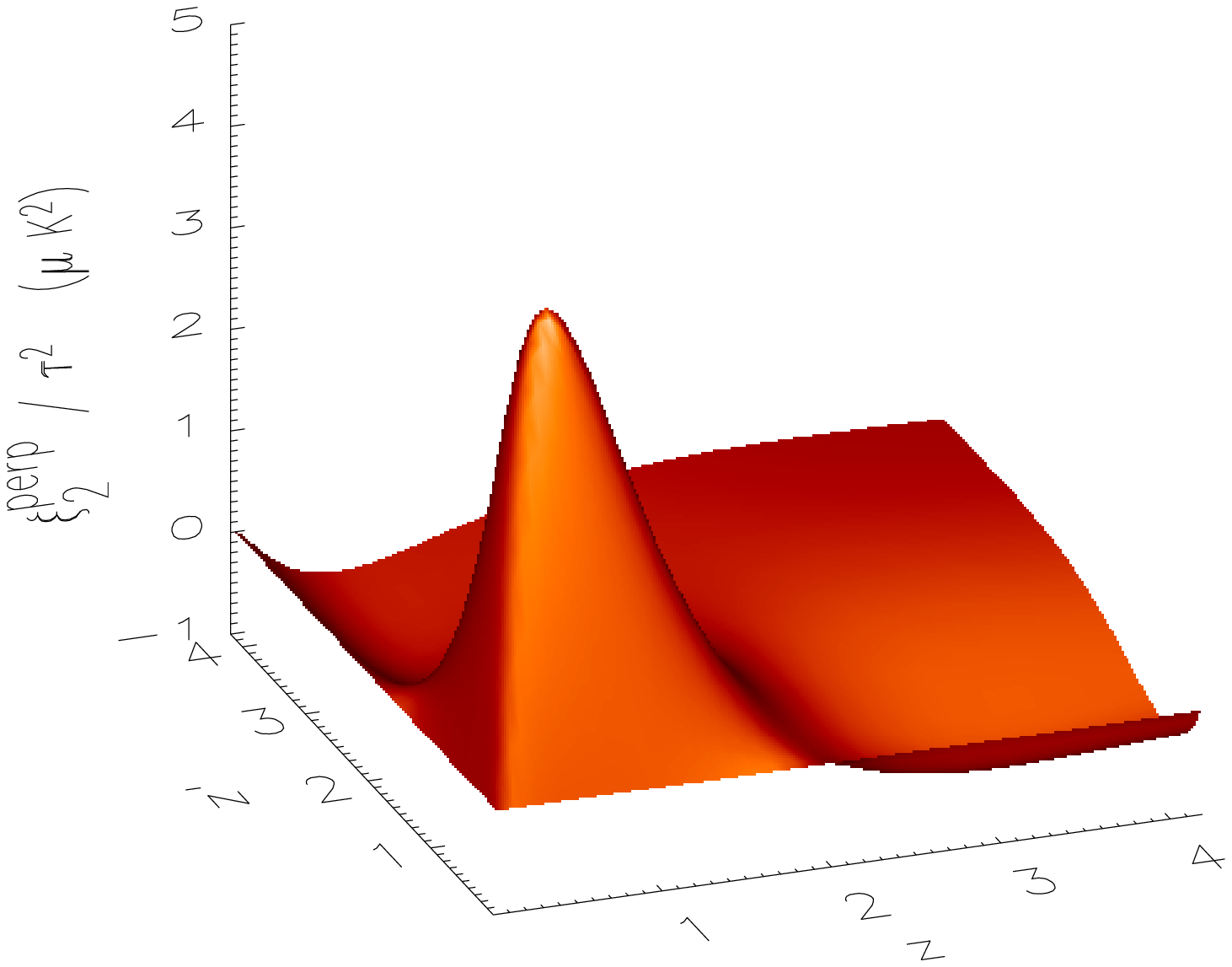}
\caption{Correlation functions $\xi_l^\perp$ for $l=2$.  In the
left panel, only the correlation with the local temperature
quadrupole has been projected out.  In the right panel, the
correlation with the polarization quadrupole has also been
removed.}
\label{fig:corrperp}
\end{figure*}

Fig.~\ref{fig:corrfuncs} shows the correlation functions $\xi_l$
for several values of $l$.  In calculating
the ISW integral and in converting from distance $r$ to
redshift $z$,  a flat Friedmann-Robertson-Walker cosmology with 
$\Omega_m=0.3$ and $\Omega_\Lambda=0.7$ was assumed.
The results are normalized to WMAP on large angular scales.
At low $l$, the correlations are extremely broad, meaning that
even a survey covering a wide range of distances will contain
few independent modes per $(l,m)$, as we will see in section
\ref{subsec:volume}.

Remote quadrupole measurements are in general correlated with 
the local CMB anisotropy and polarization. 
To assess how much independent information
is contained in the remote measurements, we need to know
how strong these correlations are.  Let us begin
by considering correlations with the locally-measured
temperature anisotropy.  If the anisotropy 
spherical harmonic coefficients are $a_{lm}$,
then the cross-correlation is
\begin{align}
\zeta_{L}(r)&\equiv\langle p_{LM}(r)a_{LM}^*\rangle
\label{eq:zeta}
\\
&=
\int_0^\infty dk\,k^2\Delta_2(k;r)\Delta_L(k;0)F_L(kr)P_\Phi(k),
\end{align}
where $\Delta_L$ is the appropriate transfer function.

If we wish to study only the information in a remote quadrupole data
set that is independent of the local anisotropy, then we should
project the mode coefficients onto the space orthogonal to
that probed by the local CMB:
\beq
p_{LM}^\perp(r)=p_{LM}(r)-\frac{\zeta_L(r)}{C_L}a_{LM},
\label{eq:ortho}
\eeq
where $C_l\equiv \langle a_{lm}a_{lm}^*\rangle$ is the usual CMB
angular power spectrum.

We can define
a correlation function $\zeta_L^E$
and perform a similar projection to remove the portion of the signal
that is correlated with the CMB polarization anisotropy coefficients
$a_{lm}^{E}$. (There is no significant correlation with
the $B$-type CMB polarization.)
In performing
this
projection, it is important to include the effects of reionization,
as most of the low-$l$ polarization signal comes from post-reionization
scattering.

Fig.~\ref{fig:corrperp} 
shows the correlation functions $\xi^\perp_l$ corresponding
to the projected coefficients $p_{lm}^\perp$ for $l=2$.
As we will see, the difference
between $\xi_l$ and $\xi_l^\perp$ decreases at higher $l$'s.
In performing the projection, the Universe
was assumed to have completely reionized at $z=11$, but the
results do not depend strongly on the details of reionization,
unless there was considerable patchiness on large scales.

\begin{figure}
\includegraphics[width=3in]{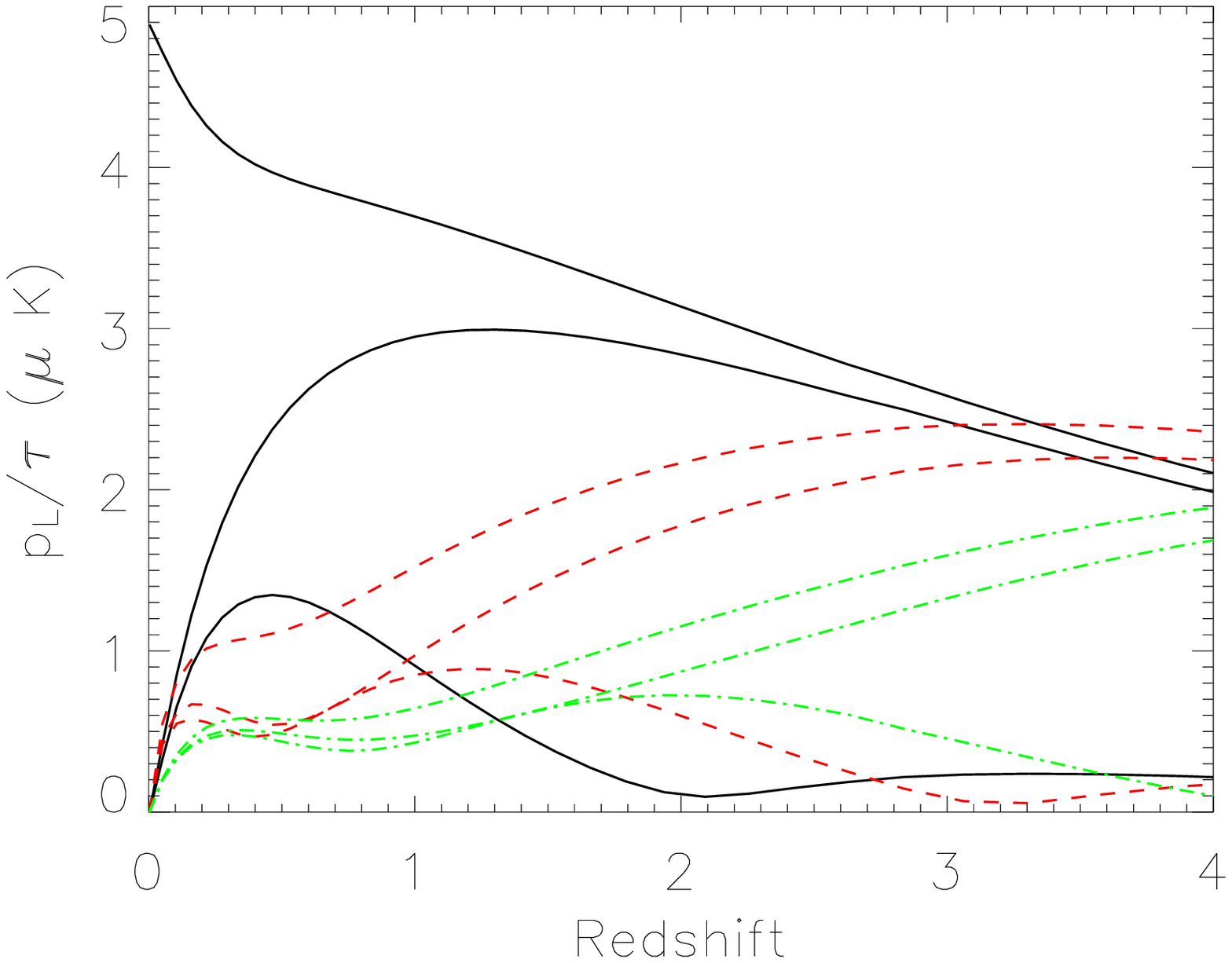}

\includegraphics[width=3in]{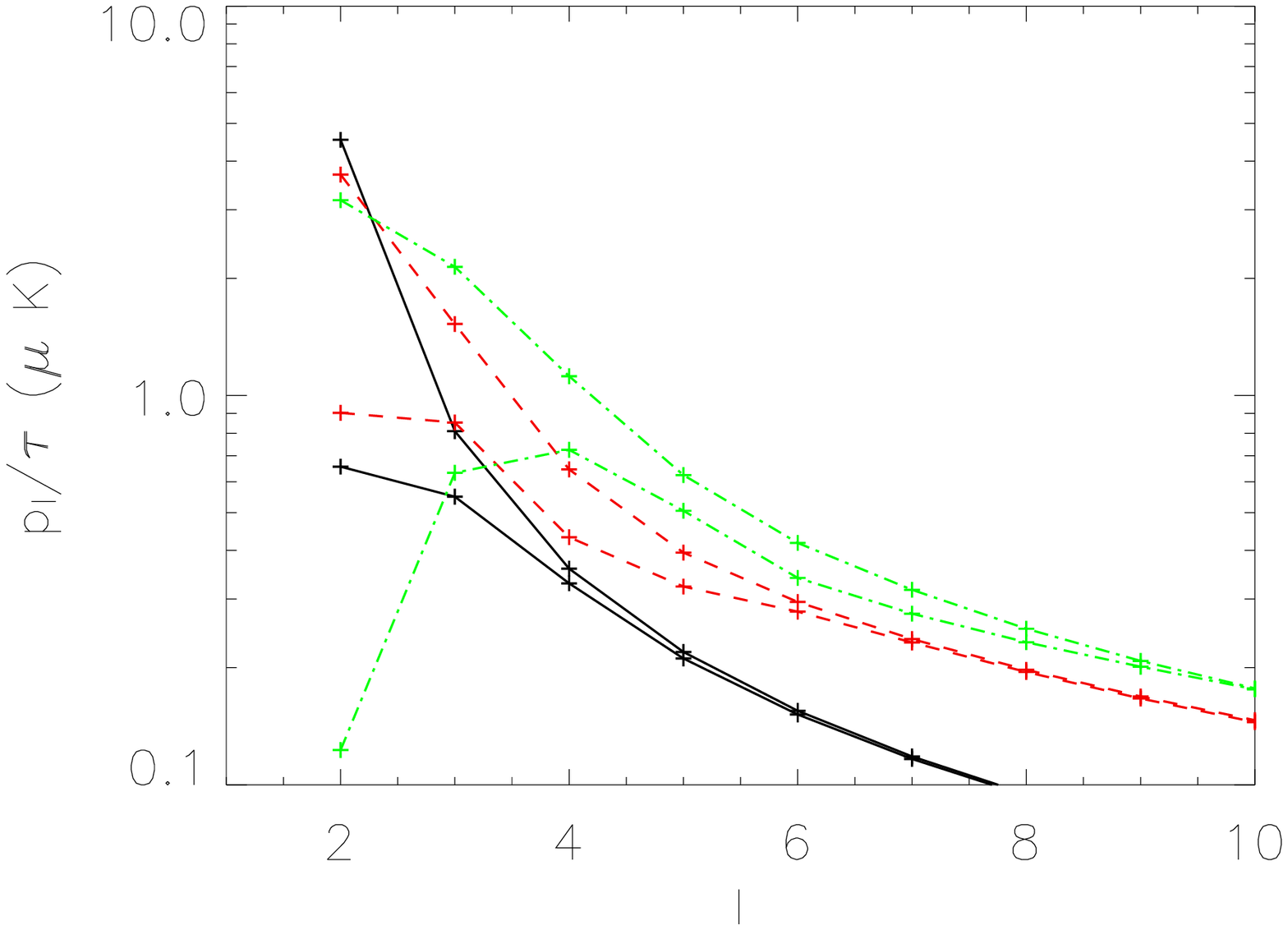}

\caption{The r.m.s.\ signal per multipole.  The left panel
shows the signal in modes $l=2$ (solid, black), 3 (dashed, red),
and 4 (dot-dashed, green) as a function of redshift.  The bumps
at $z\lesssim 0.5$ are due to the ISW effect.  In each case, the three
curves from top to bottom are the total signal, the result of projecting
out the correlation with temperature quadrupole, and the result of projecting
out both temperature and polarization.  
The right panel shows the signal as a function of $l$ for redshifts
$z=0.1$ (solid, black) 1 (dashed, red), and 2 (dot-dashed, green).  
The upper curve is total signal, and the lower curve is the
result of projecting out the correlations with local temperature
and polarization.}
\label{fig:rmssignal}.
\end{figure}

The r.m.s.\ power due to all modes at a given $l$ is
\beq
p_l(r)=\sqrt{\frac{2l+1}{4\pi}\xi_l(r,r)},
\eeq
and similarly for $p_l^\perp$.  As Fig.~\ref{fig:rmssignal}
indicates, the quadrupole $l=2$ dominates the unprojected
power at low redshift,
but power shifts to smaller angular scales (higher $l$) 
as the redshift increases.  At $l=2$, the projected power
is much less than the unprojected power at all redshifts: the modes
$p_{2m}$ are strongly correlated with the local temperature
quadrupole at low $z$ and with the polarization quadrupole
at high $z$.  Modes with $l>2$ are comparatively weakly correlated
with the local signals over some redshift ranges.

All correlation functions except $\xi_2$ go to zero as $z,z'\to 0$.
This is expected: at low redshift, the remote quadrupole $p({\bf r})$
contains precisely the same information as the local quadrupole 
coefficients $a_{2m}$, so it must transform as a quadrupole
itself.  Indeed, it is straightforward to check from equation
(\ref{eq:plm}) that 
\beq
p_{lm}(r)\to N\sqrt{\frac{4\pi}{5}}a_{lm}\delta_{l2}
\qquad\mbox{as $r\to 0$}.
\label{eq:quadcorr}
\eeq

The real-space correlation
functions are easily computed from the spherical harmonic space functions.
The correlation
between remote quadrupole signals of two clusters at locations ${\bf r}_1,{\bf r}_2$
is
\beq
\langle p({\bf r}_1)p({\bf r}_2)\rangle=
\sum_{L=2}^\infty\frac{2L+1}{4\pi}\xi_L(r_1,r_2)P_L(\rhat_1\cdot\rhat_2)
\eeq
using equations (\ref{eq:ylmexpansion}) and (\ref{eq:plmcorr}) and
the spherical harmonic addition theorem.
Similarly, the correlation between a remote quadrupole measurement
$p({\bf r}_1)$ and the local CMB $(\Delta T/T)({\bf r}_2)$ is
\beq
\langle p({\bf r}_1)(\Delta T/T)(\rhat_2)\rangle=
\sum_{L=2}^\infty \frac{2L+1}{4\pi}\zeta_L(r_1)P_L(\rhat_1\cdot\rhat_2).
\eeq
Once the initial investment of calculating the $l$-space
correlation functions has been made, these formulae allow rapid
calculation of real-space correlations.

\section{Scales probed by remote quadrupole surveys}
\label{sec:surveymodes}
\subsection{Survey at a fixed redshift}

\begin{figure}
\includegraphics[width=3.2in]{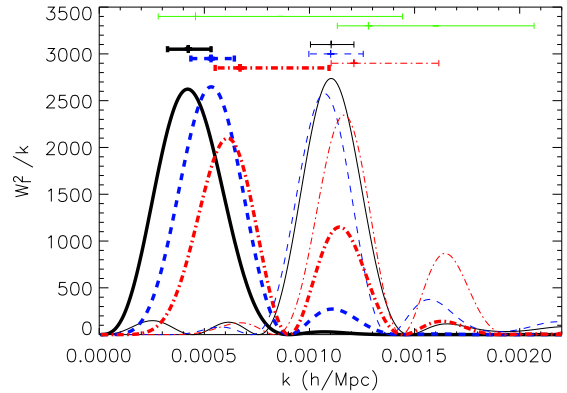}
\caption{Window functions.  The quantity $W_l^2/k$, which is
proportional to the mean-square power in multipole a given multipole, is
shown for a survey at $z=2$ for $l=2$ (solid, black), 3 (dashed, blue),
and 4 (dot-dashed, red).  In each case, the left (thicker) curve
is the window function for the total signal $p_{lm}$,
and the right (thinner) curve is the window function for $p_{lm}^\perp$.
All
functions are normalized
to integrate to one.  
The bars above
each window function show the 25th, 50th, and 75th
percentiles.  The green bars at the top of the plot indicate the
percentiles for the $l=2$ and $l=10$ multipoles of the local CMB anisotropy.}
\label{fig:window}
\end{figure}


We next examine the length scales probed by the various multipoles, assuming
an all-sky survey  has been used to estimate the
coefficients $p_{lm}$ at some fixed redshift.
We can write the signal as 
\beq
p_{l}^2(r)=\int dk\,k^2 P_\Phi(k)W_l^2(k; r),
\eeq 
with $W_l(k;r)=\frac{2l+1}{4\pi} \Delta_2(k;r)F_l(kr)$.
Since $P_\Phi\propto k^{-3}$, the quantity $W_l^2/k$ is
proportional to the power per wavenumber interval $dk$.  Similarly,
we can define a window function for the quantity $p_l^\perp$ that results
from projecting out the part of the signal that is correlated with
the local CMB.

The first few window functions are 
shown in Fig.~\ref{fig:window} for $z=2$.  Window functions
corresponding to both $p_l$ and $p_l^\perp$ are shown.
The range of scales
probed by the various window functions are indicated with horizontal
error bars, and for comparison the ranges corresponding to the 
local CMB power spectrum $C_2$ and $C_{10}$ are also indicated.
Because of the ISW effect, the local CMB window functions are quite
broad.  

The first few unprojected modes probe scales as large as the CMB
quadrupole but with narrower window functions.  As noted earlier,
these are significantly correlated with the local CMB polarization
multipoles.  Nonetheless, considering the likelihood that large-angle
CMB polarization multipoles may be contaminated by foregrounds
or systematic errors,
the unprojected modes will still provide valuable new
information on the largest-scale perturbations in the Universe, or
at least test our understanding of large-angle polarization data.
(The remote quadrupole survey will of course be susceptible
to systematic errors and foregrounds as well, but the susceptibility
will be different from that of the local polarization multipoles.)

The projected modes probe smaller scales, but they are still
in the gigaparsec range, comparable to the first 10 or so CMB
multipoles, and in some cases have narrow window functions.
In practice, the projected modes at $l=2$ ($p_2^\perp$) are
unlikely to be reliably measured, because the correlations
are so strong, but projected modes with $l\ge 3$ will allow
us to probe these large scales.

\begin{figure*}
\includegraphics[width=5in]{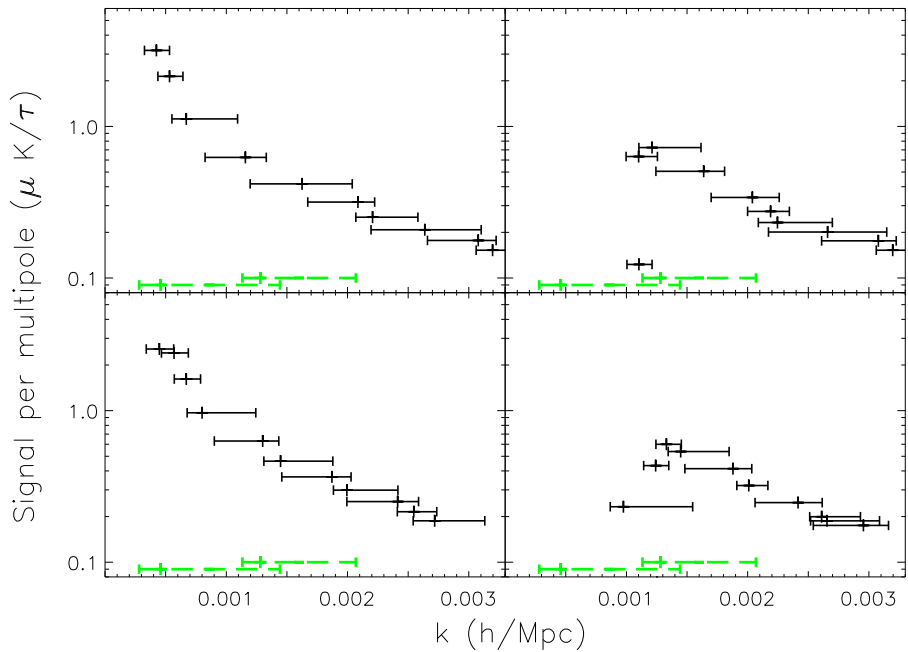}
\caption{The r.m.s.\ signal  vs.\ effective length
scale.  The upper plots are for a survey at $z=2$, and the
lower plots are for $z=3$.  On the left the 
total signal $p_{l}$ is shown, and on the right is the projected
signal $p_l^\perp$.
The horizontal error bars indicate 25th, 50th,
75th percentile contributions to the signal. Multipole
$l$ increases from left to right within each plot.
The dashed green bars at the bottom indicate the window functions
for the local CMB anisotropy at $l=2$ and $l=10$; the vertical
position of these bars is arbitrary.}
\label{fig:sigvsscale}
\end{figure*}

Fig.~\ref{fig:sigvsscale} shows the r.m.s.\ power $p_l$ and $p_l^\perp$
per multipole, plotted against the effective scale for each
multipole.  
In interpreting this plot, bear in mind that each point represents
the r.m.s.\ signal from all $2l+1$ modes at a given $l$.

In order to measure the quantities $p_l^\perp$, in principle
we need an all-sky cluster survey, knowledge of the local
CMB anisotropy and polarization spherical
harmonic coefficients, and knowledge of the
correlation functions $\zeta_l(r)$ and $\zeta_l^E(r)$ in order to project out
the local contribution.  In practice, of course,
difficulties are likely with all of these.  Section \ref{sec:discussion}
contains some discussion of how to mitigate these problems.
For the moment, observe that information on large physical
scales is found at large angular scales.  We must survey a large fraction
of the sky if we want to address the puzzles in the large-scale CMB
with this technique.  However, note that at redshifts $z=2$-3
the 
signal drops fairly rapidly as a function of $l$.  This is good
news: it means that a relatively sparse survey can measure 
the low-$l$ modes without excessive contamination from small angular
scales.

\subsection{Volume-limited survey}
\label{subsec:volume}

In the previous subsection, we considered surveys at a fixed redshift.
We now imagine a volume-limited survey out to some maximum
redshift $z_{\rm max}$.  Let us continue to assume an all-sky
survey, so that it is natural to think of the survey in spherical
harmonic space.  In this case,
our survey provides estimates of each of the functions $p_{lm}(r)$
at multiple values of $r$.

For each $l$, we can enumerate a list of signal strength
eigenmodes $\psi_{nl}$ that are solutions to
\beq
\int_0^{z_{\rm max}}\xi_l(r,r')\psi_{nl}(r')\,r'^2\,dr'=
\lambda_{nl}\psi_{nl}(r).
\eeq
The mode functions $\psi_{nl}(r)\ \Ytwo_{lm}(\rhat)$ form an orthonormal
basis, which we can use to express
the signal $p({\bf r})$.  
The mean-square signal in each mode is
the eigenvalue $\lambda_{nl}$, so these modes provide
a useful guide to tell us where the signal is strong.

\begin{figure*}
\includegraphics[width=5in]{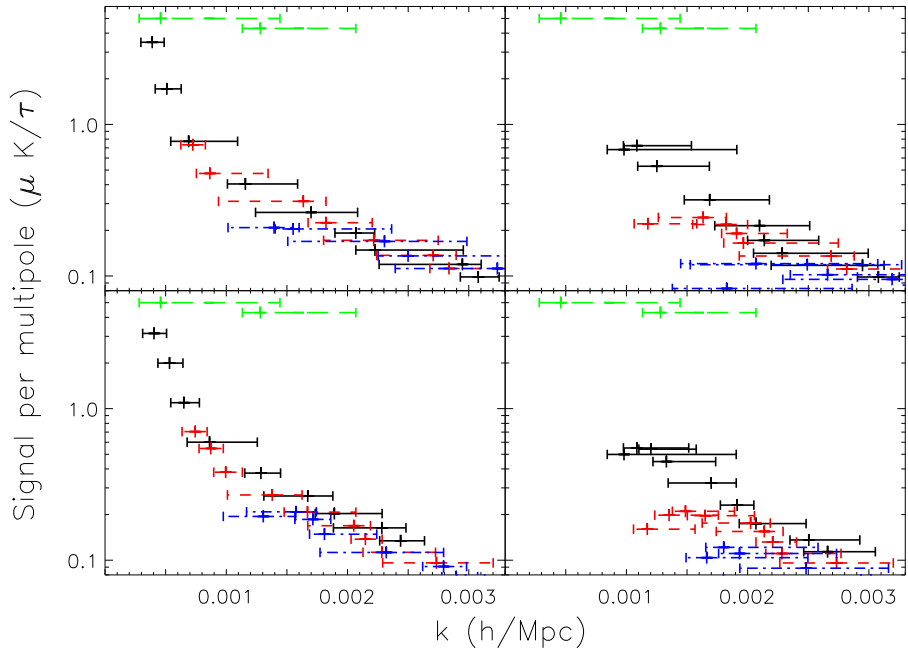}
\caption{The r.m.s.\ signal  vs.\ effective length
scale.  The upper panels show the results of a survey
out to a maximum redshift of 2, and the lower panels are for
a maximum redshift of 3.  The total signal $p_l$ is plotted
on the left, and the projected signal $p_l^\perp$ is on the right.
The horizontal axis indicates the length scales of each
mode as in Fig.~\ref{fig:sigvsscale}.
Solid black bars are the highest signal-strength eigenmode
as a function of $l$.  The dashed red bars are the
second mode for each $l$, and the dot-dashed blue bars are the
third mode.  Multipole $l$ increases from left to right
within each category.
The long-dashed green bars at the top indicate the window functions
for the local CMB anisotropy at $l=2$ and $l=10$; the vertical
position of these bars is arbitrary.}
\label{fig:sigvol}
\end{figure*}

For each
mode, we can calculate a window function and hence
assign a range of wavenumbers probed as we did for the surveys at fixed
redshift.  Results are illustrated in Fig.~\ref{fig:sigvol}.

As one would expect, the mode with highest signal at
each $l$ corresponds
to a simple weighted average of $p_{lm}$ as a function of $r$
with positive weight everywhere.  The next mode 
is essentially a difference between low- and high-redshift
signals, and subsequent modes contain more radial oscillations.
As Fig.~\ref{fig:sigvol} indicates, only the first
couple of modes are likely to be measurable at any given $(l,m)$.
Once again, the $l=2$ modes are strongly correlated with 
the local CMB polarization.  Assuming the large-angle CMB polarization
has been well measured, they provide relatively little new
information; modes with
$l\ge 3$ are the richest source of independent data on
large-scale perturbations.  On the other hand, if the first few 
CMB polarization multipoles are uncertain due to foregrounds or systematic
errors, then the $l=2$ modes of a remote quadrupole survey may help to fill
in this gap.

A comparison of Figs.~\ref{fig:sigvsscale} and \ref{fig:sigvol}
shows that significantly more large-scale information can be obtained
from a volume-limited survey than from a survey on a shell.  
There is, of course, an obvious
price to pay: 
many more clusters must be observed to estimate all these modes.

\section{Discussion}
\label{sec:discussion}

We have seen that 
an all-sky survey can probe the gigaparsec-scale Universe, measuring
fluctuation modes that are independent of the local CMB.  
In surveys at redshifts around 2-3, the large-angular scale modes
provide data on perturbations on the same length scales as the
first few CMB multipoles, but with quite narrow
window functions.

The results shown in the previous section were for an idealized survey:
in addition to full sky coverage, 
the local CMB anisotropy coefficients $a_{lm}$ and the
cross-correlations $\zeta_l(r)$, as well
as the corresponding quantities for polarization, 
were assumed to be known in order
to compute the projected signal $p_{lm}^\perp$.
We must ask what happens if these assumptions are replaced
by more realistic ones.  The most complete way to answer
these questions would be to assume a precise survey geometry
and compute the resulting Fisher matrix.  We will not perform
such a detailed analysis here; we can, however, make some general observations.

In 
a survey that covers a fraction of the sky $f_{\rm sky}$,
only band powers with width $\Delta l\sim
f_{\rm sky}^{-1/2}$ can be recovered, not individual multipoles.
Furthermore, 
the lowest-$l$ modes
cannot be recovered at all.  For the goal of probing
the largest scales, therefore, large sky coverage
is essential independent of the choice of redshift.  
A survey with $f_{\rm sky}=0.1$, for instance (4000 square degrees)
would be able to recover only a single mode in the $l=2$-$3$ band.

On the other
hand, the power drops fairly
rapidly as a function of $l$, so contamination
of the low-$l$ modes from high-$l$ power is modest.  In other
words, in order to probe large scales, we should
survey as much sky as possible, but the survey can be sparse.

Next, let us consider uncertainties in projecting out the local CMB
contribution (i.e., going from $p_{lm}$ to $p_{lm}^\perp$).  
For all $l>2$, this projection is 
subdominant to the primary signal over some
range of redshifts, so independent information 
should be obtainable from these modes.

The $l=2$ modes
are a different matter, as the correlations are extremely strong
there.  As Fig.~\ref{fig:rmssignal} indicates, the projected
coefficients $p_{2m}^\perp$ are much smaller than the unprojected
coefficients $p_{2m}$ at all redshifts.  At low redshift, the culprit
is the temperature quadrupole, while at high redshift $p_{2m}$ becomes
very strongly correlated with the polarization quadrupole.
To put the situation pessimistically, accurate extraction of
$p_{2m}^\perp(r)$ may never be feasible.  To extract information about
large-scale perturbations that is independent of the local CMB, we
will look to modes $p_{lm}$ with $l\ge 3$ (or, in the case of a
partial-sky survey, by modes that cover the largest available angular
scales but are orthogonal to the quadrupole).

A more optimistic interpretation is that
measurement of $p_{2m}$ at a couple of different redshifts
can allow us to determine the {\it local} CMB temperature
and polarization quadrupole coefficients (that is, the 5 coefficients
$a_{2m}$ and the 5 coefficients $a_{2m}^E$).  Since direct 
measurements of these coefficients may be contaminated by foregrounds
or systematic errors (especially in the case of polarization), such an
independent determination of these coefficients will be important
in assessing the significance of the large-scale anomalies in 
the CMB.  Furthermore, by measuring $p_{2m}$ as a function of redshift,
we may be able to test our theoretical predictions of the cross-correlation
functions $\zeta_2,\zeta_2^E$, thus providing a probe of the recent ISW effect.

\begin{figure}
\includegraphics[width=3in]{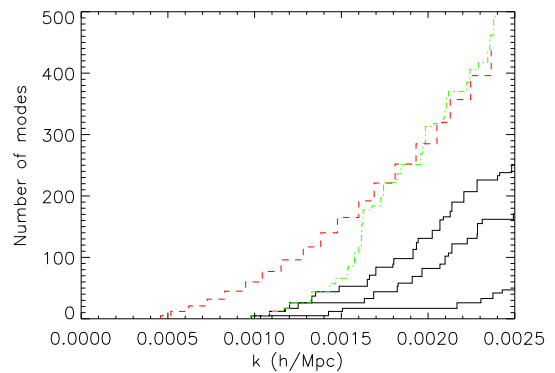}
\caption{The cumulative number of modes as a function
of scale.  From bottom to top, the solid curves show predictions
for surveys with $z_{\rm max}=1,2,3$.  The red dashed curve
shows the number of modes probed by the local CMB anisotropy,
and the green dotted curve shows the results of a hypothetical
remote quadrupole all-sky survey out to $z=\infty$.}
\label{fig:modecount}
\end{figure}

A common question is whether a remote quadrupole survey can ``beat cosmic
variance.''  The answer depends on precisely what we mean by this
phrase.  Fig.~\ref{fig:modecount} provides one possible answer.
The figure shows the cumulative number of independent
modes on scales larger than a given value for ideal all-sky surveys
out to a specified redshift.  All of the signal eigenmodes are
included in this count.  The number of modes contained in the all-sky
CMB temperature anisotropy data (without polarization) is shown for
comparison.  Although a survey that went all the way out to $z=\infty$
would ``beat'' the local CMB, realistic surveys never do.  
The amount of new information can be comparable, however, on some
scales.  In particular, the number of new modes obtainable
by a remote quadrupole survey in the range $k\sim 0.002h/$Mpc
is about the same as that contained in the local CMB (because
the slopes of the cumulative curves in Fig.~\ref{fig:modecount}
are about the same there).
Considering the
unsettled state of our understanding of gigaparsec-scale perturbations
and the hints that something surprising may be going on there, it is
clear that there is valuable information to be gained.

This article has focused primarily on the largest-scale information
contained in remote cluster surveys.  The formalism described here is
also useful for surveys designed to probe the ISW effect
\cite{baumanncooray,CoorayBaumann,CHB,seto}.
Such surveys provide a powerful
probe of the recent growth of structure
and hence may shed light on the nature
of dark energy and the growth of structure.  Because the ISW effect
is most important at low redshift (see Fig.\ \ref{fig:rmssignal}),
such a survey will be quite different from those considered here:
the best approach appears to be a denser survey of a smaller area
of the sky at low redshift.
In planning a survey to probe the ISW effect, 
it will be important to quantify the number
of independent modes that can be probed.  The detailed answer 
will depend on the precise locations of the clusters to be surveyed,
but a simple estimate obtained by
counting
modes in spherical harmonic space and scaling by $f_{\rm sky}^{-1/2}$
should provide valuable guidance.  

Clusters are of course not randomly distributed ``test particles'':
they are overdensities.  One might worry that this would lead to
biases in the modes recovered from such a survey.  A remote quadrupole
survey (even a small-scale one optimized for characterizing the ISW)
primarily probes scales of several hundred Mpc or more, which is
considerably larger than the scale associated with the formation of
individual clusters.  One would therefore not expect significant bias
due to the locations of individual clusters.  On the other hand, the
modes recovered from such a survey would presumably be correlated with
tracers of large-scale structure on hundred-Mpc scales.  In analyzing
the results of such a future survey, one would want to characterize
those correlations, presumably via $N$-body simulations.  For the
gigaparsec-scale surveys that are the primary focus of this paper,
of course, clusters can be taken as randomly-distributed test particles.

In a detailed Fisher-matrix
analysis of a potential survey, the real-space covariance matrix
$\langle p({\bf r}_i)p({\bf r}_j)\rangle$ will be needed, as will
the correlation with the local CMB.  The
formalism in this paper provides a useful way to compute these quantities.
The full covariance matrix can be computed with and without
the ISW effect, and Fisher matrix estimates of the errors with which
ISW parameters can be reconstructed can be computed quickly
and easily for any desired survey geometry.

A survey of the sort considered here will surely be a daunting task.
The signals are sub-microkelvin, and there is unfortunately no
shortage of confusing signals.  Some signals (diffuse Galactic
foregrounds and the kinematic signal due to the cluster's peculiar
velocity) can be
distinguished by their spectral signature, assuming a multifrequency
survey, but detailed information on the spectral and spatial properties
of polarized foregrounds will be necessary.
The three-year WMAP data have advanced the state of knowledge
in this area considerably \cite{wmap3pol}, and further information 
will come from
the Planck satellite \cite{tauber} as well as ground-based experiments.

Of potentially greater concern is the intrinsic CMB
polarization (both due to last scattering and reionization), which
will be lensed by the cluster itself.  In order for the remote
quadrupole survey to be detectable, we will probably need detailed
knowledge of the cluster optical depth as a function of position on
the sky, and possibly the projected mass density as well.  With this
information, we can construct a template for the remote quadrupole
signal and use it to fit for the two parameters that determine the
remote quadrupole at that cluster [$Q$ and $U$ or equivalently the
real and imaginary parts of $p({\bf r})$].  Since the background
polarization is not expected to be spatially correlated with this
template, this will help in separating the signal from the confusing
background.  In the near future, Sunyaev-Zel'dovich surveys will
provide a wealth of detailed cluster data \cite{SPT,ACT}, so there is
reason to hope that such an approach may soon be feasible.

In any remote quadrupole survey, assessment of the errors
will be crucial.  For instance,  errors in determining the optical
depths $\tau$ of the clusters can induce spurious signals.
Just as in the case of CMB polarization maps, a valuable diagnostic
can be obtained by considering the decomposition of the
data into $E$ and $B$ modes \cite{eb1,eb2}.  At any fixed redshift $z$,
the remote quadrupole data consist of a spin-2 field on the sphere
that is derived from a scalar perturbation (assuming that
primordial tensor perturbations can be neglected).  As noted
in Section \ref{sec:formalism}, the true signal ---
everything calculated in this paper --- 
should therefore consist only of $E$ modes, precisely as in the
case of scalar perturbations in the CMB.  Noise and systematic errors, on the
other hand, are likely to populate E and B equally \cite{zaleb}.
When analyzing results of an actual survey, therefore, the B
modes can be monitored to determine the errors.
In practice, for a partial-sky survey with sparse sampling,
there will be significant $E$-$B$ mixing \cite{LCT,bunn,bunnerratum,bunnetal},
but this technique should still provide a valuable check.


\paragraph*{Acknowledgments:}
This work was supported by NSF Grants 0233969 and 0507395 
and by a Cottrell Award
from the Research Corporation.  I thank 
Max Tegmark and the MIT physics department
for their hospitality during the completion of this work,
and an anonymous referee for helpful comments.

\appendix

\section{}

In this section we derive some identities involving spherical
harmonics, coordinate transforms, and 3-$j$ symbols.

\subsection{Rotation matrices and spin-2 spherical harmonics.}
For a given cluster location ${\bf r}$, we adopt a primed
coordinate system obtained by rotating the $z$ axis until
it points in the direction $\rhat$.  
Let $R$ be the
rotation that relates the two coordinate systems.
The Euler angles
associated with this rotation are $(\phi_{\rhat},\theta_{\rhat},0)$,
using the same conventions as \cite{LCT}.
The spherical harmonic $Y_{22}$ in the primed coordinate
system can be expressed in the unprimed coordinates as
\beq
Y_{22}(\khat')=
\sum_{m=-2}^2 D^2_{m2}(R)Y_{2m}(\khat),
\eeq
where $D^l_{mm'}(R)$ is the Wigner matrix for the rotation $R$.
The Wigner matrices can be expressed in terms of spin-$s$ spherical
harmonics \cite{LCT}:
\beq
D^l_{-ms}(\phi,\theta,-\psi)=(-1)^m\sqrt{\frac{4\pi}{2l+1}}
\ {}_sY_{lm}(\theta,\phi)e^{is\psi}.
\eeq
The result is
\beq
Y_{22}(\khat')
=\sqrt{\frac{4\pi}{5}}\sum_m (-1)^m\Ytwo_{2-m}(\hat{\bf r})Y_{2m}(\khat).
\label{Y22equation}
\eeq

\subsection{Integrals over spherical harmonics.}
The derivation in Section \ref{sec:formalism} contains
an integral 
\beq
I\equiv\int d^2\rhat\ \Ytwo_{2-m}(\rhat)\ 
\Ytwo_{LM}^*(\rhat)e^{i{\bf k}\cdot{\bf r}}.
\label{Idef}
\eeq
To evaluate this
integral, we expand the exponential in spherical harmonics to get
\begin{widetext}
\beq
I=4\pi\sum_{\lambda,\mu} i^\lambda j_\lambda(kr) 
Y_{\lambda\mu}(\khat)
\int d^2\rhat\,
Y_{\lambda\mu}^*(\rhat)
\ \Ytwo_{2-m}(\rhat)\ 
\Ytwo_{LM}^*(\rhat).
\eeq
Next, we want to evaluate the integral
over the three spherical harmonics.  Using
the identity ${}_sY_{lm}^*=(-1)^{m+s}\,{}_{-s}Y_{l-m}$, the
integral we need can be written in the form
\beq
J\equiv \int d^2\rhat\ 
\Ytwo_{l_1m_1}(\rhat)\ {}_{-2}Y_{l_2m_2}(\rhat)Y_{l_3m_3}(\rhat).
\eeq
Equation (B3) of \cite{LCT} tells us how to express
all of the spherical harmonics in terms of $D$-matrices:
\beq
J=\sqrt{\frac{(2l_1+1)(2l_2+1)(2l_3+1)}{(4\pi)^3}}\int d^2\rhat\ D^{l_1}_{-m_12}
D^{l_2}_{-m_22}D^{l_3}_{-m_30}.
\eeq
Here the $D$-matrices can be evaluated for any rotation $R(\phi,\theta,\psi)$
with the first two Euler angles being the spherical coordinates of $\rhat$.
Since the integrand doesn't depend on the third Euler angle $\psi$, we 
can replace $\int d^2\rhat$ with $\frac{1}{2\pi}\int d^3R$, an integral
over the entire rotation group.  Zare \cite{zare} (p. 103) gives this integral
in terms of $3j$ symbols:
\beq
J=\sqrt{\frac{(2l_1+1)(2l_2+1)(2l_3+1)}{4\pi}}
\begin{pmatrix}
l_1 & l_2 & l_3\\ 
-m_1 & -m_2 & -m_3
\end{pmatrix}
\begin{pmatrix}
l_1 & l_2 & l_3\\ 2 & -2 & 0
\end{pmatrix}.
\label{eq:int3j}
\eeq
So we can write the integral in equation (\ref{Idef})
as
\beq
I=\sum_{\lambda,\mu}i^\lambda(-1)^m j_\lambda(kr)Y_{\lambda\mu}(\khat)
\sqrt{20\pi(2\lambda+1)(2L+1)}
\begin{pmatrix}
2 & L & \lambda\\ 2 & -2 & 0
\end{pmatrix}
\begin{pmatrix}
2 & L & \lambda\\ m & M & \mu
\end{pmatrix}.
\eeq

We can use this result to write equation (\ref{blmequation})
as
\beq
p_{LM}(r)=4\pi N
\int d^3k\,\Delta_2(k;r)\delta_\Phi({\bf k})
\sum_{\lambda}i^\lambda \sqrt{(2L+1)(2\lambda+1)}
j_\lambda(kr)
\begin{pmatrix}
2 & L & \lambda\\ 2 & -2 & 0\end{pmatrix}
K(\khat),
\label{eq:uglyblm}
\eeq
where
\beq 
K(\khat)=\sum_{m,\mu}
\begin{pmatrix}2 & L & \lambda\\ m & M & \mu\end{pmatrix}
Y_{\lambda\mu}(\khat)Y_{2m}(\khat).
\eeq
Expand $K$ in spherical harmonics: $K(\khat)=\sum_{l_0m_0}K_{l_0m_0}
Y_{l_0m_0}(\khat)$.
The coefficients 
are
\begin{align}
K_{l_0m_0}&=
\sum_{m,\mu}
\begin{pmatrix}2 & L & \lambda\\ m & M & \mu\end{pmatrix}
\int d^2\khat\, Y_{2m}(\khat)
Y_{\lambda\mu}(\khat)Y_{l_0m_0}^*(\khat)\\
&= 
(-1)^{m_0}\sum_{m,\mu}
\begin{pmatrix}2 & L & \lambda\\ m & M & \mu\end{pmatrix}
\sqrt{\frac{5(2l_0+1)(2\lambda+1)}{4\pi}}
\begin{pmatrix}2 & l_0 & \lambda\\ 0 & 0 & 0\end{pmatrix}
\begin{pmatrix}2 & l_0 & \lambda\\ m & -m_0 & \mu\end{pmatrix}\\
&=\frac{(-1)^{m_0}}{2L+1}
\sqrt{\frac{5(2l_0+1)(2\lambda+1)}{4\pi}}
\begin{pmatrix}2 & l_0 & \lambda\\ 0 & 0 & 0\end{pmatrix}
\delta_{Ll_0}\delta_{M,-m_0},
\end{align}
using equation (3.119) in \cite{zare} to integrate
the product of three spherical harmonics, and then 
using the orthogonality of the 3-$j$ symbols [equation (2.32) in \cite{zare}].
So $K$ has only one term in its spherical harmonic expansion:
$K(\khat)=K_{L-M} Y_{L-M}(\khat)$.  Substituting this into equation
(\ref{eq:uglyblm}), we get
\beq
p_{LM}(r)=
N
\sqrt{20\pi}
\sum_\lambda i^\lambda(2\lambda+1)
\begin{pmatrix}2 & L & \lambda\\ 2 & -2 & 0\end{pmatrix}
\begin{pmatrix}2 & L & \lambda\\ 0 & 0 & 0\end{pmatrix}
\int d^3k\,\Delta_2(k;r)\delta_\Phi({\bf k})j_\lambda(kr)Y_{LM}^*(\khat).
\label{eq:blmnotugly}
\eeq
\end{widetext}
 The 3-$j$ symbols vanish whenever the triangle inequality
is not satisfied, so $\lambda$ must be between $L-2$ and $L+2$.
Furthermore, $\left(\begin{smallmatrix}2 & L & \lambda\\ 0 & 0 & 0
\end{smallmatrix}\right)=0$ when $2+L+\lambda$ is odd.
So the sum above contains
only three terms: $\lambda=L-2,L,L+2$.

\bibliography{remote}

\begin{thebibliography}{45}
\expandafter\ifx\csname natexlab\endcsname\relax\def\natexlab#1{#1}\fi
\expandafter\ifx\csname bibnamefont\endcsname\relax
  \def\bibnamefont#1{#1}\fi
\expandafter\ifx\csname bibfnamefont\endcsname\relax
  \def\bibfnamefont#1{#1}\fi
\expandafter\ifx\csname citenamefont\endcsname\relax
  \def\citenamefont#1{#1}\fi
\expandafter\ifx\csname url\endcsname\relax
  \def\url#1{\texttt{#1}}\fi
\expandafter\ifx\csname urlprefix\endcsname\relax\def\urlprefix{URL }\fi
\providecommand{\bibinfo}[2]{#2}
\providecommand{\eprint}[2][]{\url{#2}}

\bibitem[{\citenamefont{{Bennett} et~al.}(2003)\citenamefont{{Bennett},
  {Halpern}, {Hinshaw}, {Jarosik}, {Kogut}, {Limon}, {Meyer}, {Page},
  {Spergel}, {Tucker} et~al.}}]{wmapbasic}
\bibinfo{author}{\bibfnamefont{C.~L.} \bibnamefont{{Bennett}}},
  \bibinfo{author}{\bibfnamefont{M.}~\bibnamefont{{Halpern}}},
  \bibinfo{author}{\bibfnamefont{G.}~\bibnamefont{{Hinshaw}}},
  \bibinfo{author}{\bibfnamefont{N.}~\bibnamefont{{Jarosik}}},
  \bibinfo{author}{\bibfnamefont{A.}~\bibnamefont{{Kogut}}},
  \bibinfo{author}{\bibfnamefont{M.}~\bibnamefont{{Limon}}},
  \bibinfo{author}{\bibfnamefont{S.~S.} \bibnamefont{{Meyer}}},
  \bibinfo{author}{\bibfnamefont{L.}~\bibnamefont{{Page}}},
  \bibinfo{author}{\bibfnamefont{D.~N.} \bibnamefont{{Spergel}}},
  \bibinfo{author}{\bibfnamefont{G.~S.} \bibnamefont{{Tucker}}},
  \bibnamefont{et~al.}, \bibinfo{journal}{\apj Supp.}
  \textbf{\bibinfo{volume}{148}}, \bibinfo{pages}{1} (\bibinfo{year}{2003}).

\bibitem[{\citenamefont{{Hinshaw} et~al.}(2003)\citenamefont{{Hinshaw},
  {Spergel}, {Verde}, {Hill}, {Meyer}, {Barnes}, {Bennett}, {Halpern},
  {Jarosik}, {Kogut} et~al.}}]{wmapspectrum}
\bibinfo{author}{\bibfnamefont{G.}~\bibnamefont{{Hinshaw}}},
  \bibinfo{author}{\bibfnamefont{D.~N.} \bibnamefont{{Spergel}}},
  \bibinfo{author}{\bibfnamefont{L.}~\bibnamefont{{Verde}}},
  \bibinfo{author}{\bibfnamefont{R.~S.} \bibnamefont{{Hill}}},
  \bibinfo{author}{\bibfnamefont{S.~S.} \bibnamefont{{Meyer}}},
  \bibinfo{author}{\bibfnamefont{C.}~\bibnamefont{{Barnes}}},
  \bibinfo{author}{\bibfnamefont{C.~L.} \bibnamefont{{Bennett}}},
  \bibinfo{author}{\bibfnamefont{M.}~\bibnamefont{{Halpern}}},
  \bibinfo{author}{\bibfnamefont{N.}~\bibnamefont{{Jarosik}}},
  \bibinfo{author}{\bibfnamefont{A.}~\bibnamefont{{Kogut}}},
  \bibnamefont{et~al.}, \bibinfo{journal}{\apj Supp.}
  \textbf{\bibinfo{volume}{148}}, \bibinfo{pages}{135} (\bibinfo{year}{2003}).

\bibitem[{\citenamefont{{Kogut} et~al.}(2003)\citenamefont{{Kogut}, {Spergel},
  {Barnes}, {Bennett}, {Halpern}, {Hinshaw}, {Jarosik}, {Limon}, {Meyer},
  {Page} et~al.}}]{wmappol}
\bibinfo{author}{\bibfnamefont{A.}~\bibnamefont{{Kogut}}},
  \bibinfo{author}{\bibfnamefont{D.~N.} \bibnamefont{{Spergel}}},
  \bibinfo{author}{\bibfnamefont{C.}~\bibnamefont{{Barnes}}},
  \bibinfo{author}{\bibfnamefont{C.~L.} \bibnamefont{{Bennett}}},
  \bibinfo{author}{\bibfnamefont{M.}~\bibnamefont{{Halpern}}},
  \bibinfo{author}{\bibfnamefont{G.}~\bibnamefont{{Hinshaw}}},
  \bibinfo{author}{\bibfnamefont{N.}~\bibnamefont{{Jarosik}}},
  \bibinfo{author}{\bibfnamefont{M.}~\bibnamefont{{Limon}}},
  \bibinfo{author}{\bibfnamefont{S.~S.} \bibnamefont{{Meyer}}},
  \bibinfo{author}{\bibfnamefont{L.}~\bibnamefont{{Page}}},
  \bibnamefont{et~al.}, \bibinfo{journal}{\apj Supp.}
  \textbf{\bibinfo{volume}{148}}, \bibinfo{pages}{161} (\bibinfo{year}{2003}).

\bibitem[{\citenamefont{Hinshaw et~al.}(2006)\citenamefont{Hinshaw, Nolta,
  Bennett, Bean, Dor\'e, Greason, Halpern, Hill, Jarosik, Kogut
  et~al.}}]{wmap3basic}
\bibinfo{author}{\bibfnamefont{G.}~\bibnamefont{Hinshaw}},
  \bibinfo{author}{\bibfnamefont{M.}~\bibnamefont{Nolta}},
  \bibinfo{author}{\bibfnamefont{C.}~\bibnamefont{Bennett}},
  \bibinfo{author}{\bibfnamefont{R.}~\bibnamefont{Bean}},
  \bibinfo{author}{\bibfnamefont{O.}~\bibnamefont{Dor\'e}},
  \bibinfo{author}{\bibfnamefont{M.}~\bibnamefont{Greason}},
  \bibinfo{author}{\bibfnamefont{M.}~\bibnamefont{Halpern}},
  \bibinfo{author}{\bibfnamefont{R.}~\bibnamefont{Hill}},
  \bibinfo{author}{\bibfnamefont{N.}~\bibnamefont{Jarosik}},
  \bibinfo{author}{\bibfnamefont{A.}~\bibnamefont{Kogut}},
  \bibnamefont{et~al.}, \bibinfo{journal}{ArXiv Astrophysics e-prints}
  (\bibinfo{year}{2006}), \eprint{arXiv:astro-ph/0603451}.

\bibitem[{\citenamefont{Page et~al.}(2006)\citenamefont{Page, Hinshaw, Komatsu,
  Nolta, Spergel, Bennett, Barnes, Bean, Dor\'e, Halpern et~al.}}]{wmap3pol}
\bibinfo{author}{\bibfnamefont{L.}~\bibnamefont{Page}},
  \bibinfo{author}{\bibfnamefont{G.}~\bibnamefont{Hinshaw}},
  \bibinfo{author}{\bibfnamefont{E.}~\bibnamefont{Komatsu}},
  \bibinfo{author}{\bibfnamefont{M.}~\bibnamefont{Nolta}},
  \bibinfo{author}{\bibfnamefont{D.}~\bibnamefont{Spergel}},
  \bibinfo{author}{\bibfnamefont{C.}~\bibnamefont{Bennett}},
  \bibinfo{author}{\bibfnamefont{C.}~\bibnamefont{Barnes}},
  \bibinfo{author}{\bibfnamefont{R.}~\bibnamefont{Bean}},
  \bibinfo{author}{\bibfnamefont{O.}~\bibnamefont{Dor\'e}},
  \bibinfo{author}{\bibfnamefont{M.}~\bibnamefont{Halpern}},
  \bibnamefont{et~al.}, \bibinfo{journal}{ArXiv Astrophysics e-prints}
  (\bibinfo{year}{2006}), \eprint{arXiv:astro-ph/0603450}.

\bibitem[{\citenamefont{Spergel et~al.}(2006)\citenamefont{Spergel, Bean,
  Dor\'e, Nolta, Bennett, Hinshaw, Jarosik, Komatsu, Page, Peiris
  et~al.}}]{wmap3imp}
\bibinfo{author}{\bibfnamefont{D.}~\bibnamefont{Spergel}},
  \bibinfo{author}{\bibfnamefont{R.}~\bibnamefont{Bean}},
  \bibinfo{author}{\bibfnamefont{O.}~\bibnamefont{Dor\'e}},
  \bibinfo{author}{\bibfnamefont{M.}~\bibnamefont{Nolta}},
  \bibinfo{author}{\bibfnamefont{C.}~\bibnamefont{Bennett}},
  \bibinfo{author}{\bibfnamefont{G.}~\bibnamefont{Hinshaw}},
  \bibinfo{author}{\bibfnamefont{N.}~\bibnamefont{Jarosik}},
  \bibinfo{author}{\bibfnamefont{E.}~\bibnamefont{Komatsu}},
  \bibinfo{author}{\bibfnamefont{L.}~\bibnamefont{Page}},
  \bibinfo{author}{\bibfnamefont{H.}~\bibnamefont{Peiris}},
  \bibnamefont{et~al.}, \bibinfo{journal}{ArXiv Astrophysics e-prints}
  (\bibinfo{year}{2006}), \eprint{arXiv:astro-ph/0603449}.

\bibitem[{\citenamefont{{Bennett} et~al.}(1996)\citenamefont{{Bennett},
  {Banday}, {Gorski}, {Hinshaw}, {Jackson}, {Keegstra}, {Kogut}, {Smoot},
  {Wilkinson}, and {Wright}}}]{bennettcobe}
\bibinfo{author}{\bibfnamefont{C.~L.} \bibnamefont{{Bennett}}},
  \bibinfo{author}{\bibfnamefont{A.~J.} \bibnamefont{{Banday}}},
  \bibinfo{author}{\bibfnamefont{K.~M.} \bibnamefont{{Gorski}}},
  \bibinfo{author}{\bibfnamefont{G.}~\bibnamefont{{Hinshaw}}},
  \bibinfo{author}{\bibfnamefont{P.}~\bibnamefont{{Jackson}}},
  \bibinfo{author}{\bibfnamefont{P.}~\bibnamefont{{Keegstra}}},
  \bibinfo{author}{\bibfnamefont{A.}~\bibnamefont{{Kogut}}},
  \bibinfo{author}{\bibfnamefont{G.~F.} \bibnamefont{{Smoot}}},
  \bibinfo{author}{\bibfnamefont{D.~T.} \bibnamefont{{Wilkinson}}},
  \bibnamefont{and} \bibinfo{author}{\bibfnamefont{E.~L.}
  \bibnamefont{{Wright}}}, \bibinfo{journal}{\apj Lett.}
  \textbf{\bibinfo{volume}{464}}, \bibinfo{pages}{L1} (\bibinfo{year}{1996}).

\bibitem[{\citenamefont{{Gorski} et~al.}(1996)\citenamefont{{Gorski}, {Banday},
  {Bennett}, {Hinshaw}, {Kogut}, {Smoot}, and {Wright}}}]{gorskicobe}
\bibinfo{author}{\bibfnamefont{K.~M.} \bibnamefont{{Gorski}}},
  \bibinfo{author}{\bibfnamefont{A.~J.} \bibnamefont{{Banday}}},
  \bibinfo{author}{\bibfnamefont{C.~L.} \bibnamefont{{Bennett}}},
  \bibinfo{author}{\bibfnamefont{G.}~\bibnamefont{{Hinshaw}}},
  \bibinfo{author}{\bibfnamefont{A.}~\bibnamefont{{Kogut}}},
  \bibinfo{author}{\bibfnamefont{G.~F.} \bibnamefont{{Smoot}}},
  \bibnamefont{and} \bibinfo{author}{\bibfnamefont{E.~L.}
  \bibnamefont{{Wright}}}, \bibinfo{journal}{\apj Lett.}
  \textbf{\bibinfo{volume}{464}}, \bibinfo{pages}{L11} (\bibinfo{year}{1996}).

\bibitem[{\citenamefont{{Hinshaw} et~al.}(1996)\citenamefont{{Hinshaw},
  {Branday}, {Bennett}, {Gorski}, {Kogut}, {Lineweaver}, {Smoot}, and
  {Wright}}}]{hinshawcobe}
\bibinfo{author}{\bibfnamefont{G.}~\bibnamefont{{Hinshaw}}},
  \bibinfo{author}{\bibfnamefont{A.~J.} \bibnamefont{{Branday}}},
  \bibinfo{author}{\bibfnamefont{C.~L.} \bibnamefont{{Bennett}}},
  \bibinfo{author}{\bibfnamefont{K.~M.} \bibnamefont{{Gorski}}},
  \bibinfo{author}{\bibfnamefont{A.}~\bibnamefont{{Kogut}}},
  \bibinfo{author}{\bibfnamefont{C.~H.} \bibnamefont{{Lineweaver}}},
  \bibinfo{author}{\bibfnamefont{G.~F.} \bibnamefont{{Smoot}}},
  \bibnamefont{and} \bibinfo{author}{\bibfnamefont{E.~L.}
  \bibnamefont{{Wright}}}, \bibinfo{journal}{\apj Lett.}
  \textbf{\bibinfo{volume}{464}}, \bibinfo{pages}{L25} (\bibinfo{year}{1996}).

\bibitem[{\citenamefont{{de Oliveira-Costa} et~al.}(2004)\citenamefont{{de
  Oliveira-Costa}, {Tegmark}, {Zaldarriaga}, and {Hamilton}}}]{costaetal}
\bibinfo{author}{\bibfnamefont{A.}~\bibnamefont{{de Oliveira-Costa}}},
  \bibinfo{author}{\bibfnamefont{M.}~\bibnamefont{{Tegmark}}},
  \bibinfo{author}{\bibfnamefont{M.}~\bibnamefont{{Zaldarriaga}}},
  \bibnamefont{and}
  \bibinfo{author}{\bibfnamefont{A.}~\bibnamefont{{Hamilton}}},
  \bibinfo{journal}{\prd} \textbf{\bibinfo{volume}{69}},
  \bibinfo{pages}{063516} (\bibinfo{year}{2004}).

\bibitem[{\citenamefont{{Copi} et~al.}(2004)\citenamefont{{Copi}, {Huterer},
  and {Starkman}}}]{copi}
\bibinfo{author}{\bibfnamefont{C.~J.} \bibnamefont{{Copi}}},
  \bibinfo{author}{\bibfnamefont{D.}~\bibnamefont{{Huterer}}},
  \bibnamefont{and} \bibinfo{author}{\bibfnamefont{G.~D.}
  \bibnamefont{{Starkman}}}, \bibinfo{journal}{\prd}
  \textbf{\bibinfo{volume}{70}}, \bibinfo{pages}{043515}
  (\bibinfo{year}{2004}).

\bibitem[{\citenamefont{{Schwarz} et~al.}(2004)\citenamefont{{Schwarz},
  {Starkman}, {Huterer}, and {Copi}}}]{schwarz}
\bibinfo{author}{\bibfnamefont{D.~J.} \bibnamefont{{Schwarz}}},
  \bibinfo{author}{\bibfnamefont{G.~D.} \bibnamefont{{Starkman}}},
  \bibinfo{author}{\bibfnamefont{D.}~\bibnamefont{{Huterer}}},
  \bibnamefont{and} \bibinfo{author}{\bibfnamefont{C.~J.}
  \bibnamefont{{Copi}}}, \bibinfo{journal}{\prl} \textbf{\bibinfo{volume}{93}},
  \bibinfo{pages}{221301} (\bibinfo{year}{2004}).

\bibitem[{\citenamefont{{Hansen} et~al.}(2004)\citenamefont{{Hansen}, {Banday},
  and {G{\'o}rski}}}]{hansen}
\bibinfo{author}{\bibfnamefont{F.~K.} \bibnamefont{{Hansen}}},
  \bibinfo{author}{\bibfnamefont{A.~J.} \bibnamefont{{Banday}}},
  \bibnamefont{and} \bibinfo{author}{\bibfnamefont{K.~M.}
  \bibnamefont{{G{\'o}rski}}}, \bibinfo{journal}{\mnras}
  \textbf{\bibinfo{volume}{354}}, \bibinfo{pages}{641} (\bibinfo{year}{2004}).

\bibitem[{\citenamefont{{Land} and {Magueijo}}(2005)}]{land}
\bibinfo{author}{\bibfnamefont{K.}~\bibnamefont{{Land}}} \bibnamefont{and}
  \bibinfo{author}{\bibfnamefont{J.}~\bibnamefont{{Magueijo}}},
  \bibinfo{journal}{\prl} \textbf{\bibinfo{volume}{95}},
  \bibinfo{pages}{071301} (\bibinfo{year}{2005}).

\bibitem[{\citenamefont{{Bernui} et~al.}(2005)\citenamefont{{Bernui}, {Mota},
  {Reboucas}, and {Tavakol}}}]{bernui2}
\bibinfo{author}{\bibfnamefont{A.}~\bibnamefont{{Bernui}}},
  \bibinfo{author}{\bibfnamefont{B.}~\bibnamefont{{Mota}}},
  \bibinfo{author}{\bibfnamefont{M.~J.} \bibnamefont{{Reboucas}}},
  \bibnamefont{and}
  \bibinfo{author}{\bibfnamefont{R.}~\bibnamefont{{Tavakol}}},
  \bibinfo{journal}{ArXiv Astrophysics e-prints}  (\bibinfo{year}{2005}),
  \eprint{arXiv:astro-ph/0511666}.

\bibitem[{\citenamefont{{Bielewicz} et~al.}(2005)\citenamefont{{Bielewicz},
  {Eriksen}, {Banday}, {G{\'o}rski}, and {Lilje}}}]{bielewicz}
\bibinfo{author}{\bibfnamefont{P.}~\bibnamefont{{Bielewicz}}},
  \bibinfo{author}{\bibfnamefont{H.~K.} \bibnamefont{{Eriksen}}},
  \bibinfo{author}{\bibfnamefont{A.~J.} \bibnamefont{{Banday}}},
  \bibinfo{author}{\bibfnamefont{K.~M.} \bibnamefont{{G{\'o}rski}}},
  \bibnamefont{and} \bibinfo{author}{\bibfnamefont{P.~B.}
  \bibnamefont{{Lilje}}}, \bibinfo{journal}{\apj}
  \textbf{\bibinfo{volume}{635}}, \bibinfo{pages}{750} (\bibinfo{year}{2005}).

\bibitem[{\citenamefont{{Copi} et~al.}(2006)\citenamefont{{Copi}, {Huterer},
  {Schwarz}, and {Starkman}}}]{copi2}
\bibinfo{author}{\bibfnamefont{C.~J.} \bibnamefont{{Copi}}},
  \bibinfo{author}{\bibfnamefont{D.}~\bibnamefont{{Huterer}}},
  \bibinfo{author}{\bibfnamefont{D.~J.} \bibnamefont{{Schwarz}}},
  \bibnamefont{and} \bibinfo{author}{\bibfnamefont{G.~D.}
  \bibnamefont{{Starkman}}}, \bibinfo{journal}{\mnras}
  \textbf{\bibinfo{volume}{367}}, \bibinfo{pages}{79} (\bibinfo{year}{2006}),
  \eprint{astro-ph/0508047}.

\bibitem[{\citenamefont{{Bernui} et~al.}(2006)\citenamefont{{Bernui},
  {Villela}, {Wuensche}, {Leonardi}, and {Ferreira}}}]{bernui}
\bibinfo{author}{\bibfnamefont{A.}~\bibnamefont{{Bernui}}},
  \bibinfo{author}{\bibfnamefont{T.}~\bibnamefont{{Villela}}},
  \bibinfo{author}{\bibfnamefont{C.~A.} \bibnamefont{{Wuensche}}},
  \bibinfo{author}{\bibfnamefont{R.}~\bibnamefont{{Leonardi}}},
  \bibnamefont{and}
  \bibinfo{author}{\bibfnamefont{I.}~\bibnamefont{{Ferreira}}},
  \bibinfo{journal}{ArXiv Astrophysics e-prints}  (\bibinfo{year}{2006}),
  \eprint{arXiv:astro-ph/0601593}.

\bibitem[{\citenamefont{{Efstathiou}}(2003)}]{efstathiou}
\bibinfo{author}{\bibfnamefont{G.}~\bibnamefont{{Efstathiou}}},
  \bibinfo{journal}{\mnras} \textbf{\bibinfo{volume}{346}},
  \bibinfo{pages}{L26} (\bibinfo{year}{2003}).

\bibitem[{\citenamefont{{Dor{\'e}} et~al.}(2004)\citenamefont{{Dor{\'e}},
  {Holder}, and {Loeb}}}]{dore}
\bibinfo{author}{\bibfnamefont{O.}~\bibnamefont{{Dor{\'e}}}},
  \bibinfo{author}{\bibfnamefont{G.~P.} \bibnamefont{{Holder}}},
  \bibnamefont{and} \bibinfo{author}{\bibfnamefont{A.}~\bibnamefont{{Loeb}}},
  \bibinfo{journal}{\apj} \textbf{\bibinfo{volume}{612}}, \bibinfo{pages}{81}
  (\bibinfo{year}{2004}).

\bibitem[{\citenamefont{{Skordis} and {Silk}}(2004)}]{skordis}
\bibinfo{author}{\bibfnamefont{C.}~\bibnamefont{{Skordis}}} \bibnamefont{and}
  \bibinfo{author}{\bibfnamefont{J.}~\bibnamefont{{Silk}}},
  \bibinfo{journal}{ArXiv Astrophysics e-prints}  (\bibinfo{year}{2004}),
  \eprint{arXiv:astro-ph/0402474}.

\bibitem[{\citenamefont{{Sazonov} and {Sunyaev}}(1999)}]{sazonov}
\bibinfo{author}{\bibfnamefont{S.~Y.} \bibnamefont{{Sazonov}}}
  \bibnamefont{and} \bibinfo{author}{\bibfnamefont{R.~A.}
  \bibnamefont{{Sunyaev}}}, \bibinfo{journal}{\mnras}
  \textbf{\bibinfo{volume}{310}}, \bibinfo{pages}{765} (\bibinfo{year}{1999}).

\bibitem[{\citenamefont{{Kamionkowski} and {Loeb}}(1997)}]{kamionloeb}
\bibinfo{author}{\bibfnamefont{M.}~\bibnamefont{{Kamionkowski}}}
  \bibnamefont{and} \bibinfo{author}{\bibfnamefont{A.}~\bibnamefont{{Loeb}}},
  \bibinfo{journal}{\prd} \textbf{\bibinfo{volume}{56}}, \bibinfo{pages}{4511}
  (\bibinfo{year}{1997}).

\bibitem[{\citenamefont{{Seto} and {Pierpaoli}}(2005)}]{seto}
\bibinfo{author}{\bibfnamefont{N.}~\bibnamefont{{Seto}}} \bibnamefont{and}
  \bibinfo{author}{\bibfnamefont{E.}~\bibnamefont{{Pierpaoli}}},
  \bibinfo{journal}{\prl} \textbf{\bibinfo{volume}{95}},
  \bibinfo{pages}{101302} (\bibinfo{year}{2005}).

\bibitem[{\citenamefont{{Baumann} and {Cooray}}(2003)}]{baumanncooray}
\bibinfo{author}{\bibfnamefont{D.}~\bibnamefont{{Baumann}}} \bibnamefont{and}
  \bibinfo{author}{\bibfnamefont{A.}~\bibnamefont{{Cooray}}},
  \bibinfo{journal}{New Astronomy Review} \textbf{\bibinfo{volume}{47}},
  \bibinfo{pages}{839} (\bibinfo{year}{2003}).

\bibitem[{\citenamefont{{Portsmouth}}(2004)}]{portsmouth}
\bibinfo{author}{\bibfnamefont{J.}~\bibnamefont{{Portsmouth}}},
  \bibinfo{journal}{\prd} \textbf{\bibinfo{volume}{70}},
  \bibinfo{pages}{063504} (\bibinfo{year}{2004}).

\bibitem[{\citenamefont{{Cooray} and {Baumann}}(2003)}]{CoorayBaumann}
\bibinfo{author}{\bibfnamefont{A.}~\bibnamefont{{Cooray}}} \bibnamefont{and}
  \bibinfo{author}{\bibfnamefont{D.}~\bibnamefont{{Baumann}}},
  \bibinfo{journal}{\prd} \textbf{\bibinfo{volume}{67}},
  \bibinfo{pages}{063505} (\bibinfo{year}{2003}).

\bibitem[{\citenamefont{{Cooray} et~al.}(2004)\citenamefont{{Cooray},
  {Huterer}, and {Baumann}}}]{CHB}
\bibinfo{author}{\bibfnamefont{A.}~\bibnamefont{{Cooray}}},
  \bibinfo{author}{\bibfnamefont{D.}~\bibnamefont{{Huterer}}},
  \bibnamefont{and}
  \bibinfo{author}{\bibfnamefont{D.}~\bibnamefont{{Baumann}}},
  \bibinfo{journal}{\prd} \textbf{\bibinfo{volume}{69}},
  \bibinfo{pages}{027301} (\bibinfo{year}{2004}).

\bibitem[{\citenamefont{{Sachs} and {Wolfe}}(1967)}]{sachswolfe}
\bibinfo{author}{\bibfnamefont{R.~K.} \bibnamefont{{Sachs}}} \bibnamefont{and}
  \bibinfo{author}{\bibfnamefont{A.~M.} \bibnamefont{{Wolfe}}},
  \bibinfo{journal}{\apj} \textbf{\bibinfo{volume}{147}}, \bibinfo{pages}{73}
  (\bibinfo{year}{1967}).

\bibitem[{\citenamefont{{Hu} and {Dodelson}}(2002)}]{hureview}
\bibinfo{author}{\bibfnamefont{W.}~\bibnamefont{{Hu}}} \bibnamefont{and}
  \bibinfo{author}{\bibfnamefont{S.}~\bibnamefont{{Dodelson}}},
  \bibinfo{journal}{\araa} \textbf{\bibinfo{volume}{40}}, \bibinfo{pages}{171}
  (\bibinfo{year}{2002}).

\bibitem[{\citenamefont{{Challinor} et~al.}(2000)\citenamefont{{Challinor},
  {Ford}, and {Lasenby}}}]{challinorfordlasenby}
\bibinfo{author}{\bibfnamefont{A.~D.} \bibnamefont{{Challinor}}},
  \bibinfo{author}{\bibfnamefont{M.~T.} \bibnamefont{{Ford}}},
  \bibnamefont{and} \bibinfo{author}{\bibfnamefont{A.~N.}
  \bibnamefont{{Lasenby}}}, \bibinfo{journal}{\mnras}
  \textbf{\bibinfo{volume}{312}}, \bibinfo{pages}{159} (\bibinfo{year}{2000}).

\bibitem[{\citenamefont{{Shimon} et~al.}(2006)\citenamefont{{Shimon},
  {Rephaeli}, {O'Shea}, and {Norman}}}]{shimon}
\bibinfo{author}{\bibfnamefont{M.}~\bibnamefont{{Shimon}}},
  \bibinfo{author}{\bibfnamefont{Y.}~\bibnamefont{{Rephaeli}}},
  \bibinfo{author}{\bibfnamefont{B.~W.} \bibnamefont{{O'Shea}}},
  \bibnamefont{and} \bibinfo{author}{\bibfnamefont{M.~L.}
  \bibnamefont{{Norman}}}, \bibinfo{journal}{ArXiv Astrophysics e-prints}
  (\bibinfo{year}{2006}), \eprint{arXiv:astro-ph/0602528}.

\bibitem[{\citenamefont{{Liu} et~al.}(2005)\citenamefont{{Liu}, {da Silva}, and
  {Aghanim}}}]{liudasilva}
\bibinfo{author}{\bibfnamefont{G.-C.} \bibnamefont{{Liu}}},
  \bibinfo{author}{\bibfnamefont{A.}~\bibnamefont{{da Silva}}},
  \bibnamefont{and}
  \bibinfo{author}{\bibfnamefont{N.}~\bibnamefont{{Aghanim}}},
  \bibinfo{journal}{\apj} \textbf{\bibinfo{volume}{621}}, \bibinfo{pages}{15}
  (\bibinfo{year}{2005}).

\bibitem[{\citenamefont{{Padmanabhan}}(2003)}]{padmanabhan}
\bibinfo{author}{\bibfnamefont{T.}~\bibnamefont{{Padmanabhan}}},
  \bibinfo{journal}{\physrep} \textbf{\bibinfo{volume}{380}},
  \bibinfo{pages}{235} (\bibinfo{year}{2003}).

\bibitem[{\citenamefont{{Kamionkowski}
  et~al.}(1997)\citenamefont{{Kamionkowski}, {Kosowsky}, and {Stebbins}}}]{eb1}
\bibinfo{author}{\bibfnamefont{M.}~\bibnamefont{{Kamionkowski}}},
  \bibinfo{author}{\bibfnamefont{A.}~\bibnamefont{{Kosowsky}}},
  \bibnamefont{and}
  \bibinfo{author}{\bibfnamefont{A.}~\bibnamefont{{Stebbins}}},
  \bibinfo{journal}{\prd} \textbf{\bibinfo{volume}{55}}, \bibinfo{pages}{7368}
  (\bibinfo{year}{1997}).

\bibitem[{\citenamefont{{Zaldarriaga} and {Seljak}}(1997)}]{eb2}
\bibinfo{author}{\bibfnamefont{M.}~\bibnamefont{{Zaldarriaga}}}
  \bibnamefont{and} \bibinfo{author}{\bibfnamefont{U.}~\bibnamefont{{Seljak}}},
  \bibinfo{journal}{\prd} \textbf{\bibinfo{volume}{55}}, \bibinfo{pages}{1830}
  (\bibinfo{year}{1997}).

\bibitem[{\citenamefont{{Tauber}}(2004)}]{tauber}
\bibinfo{author}{\bibfnamefont{J.~A.} \bibnamefont{{Tauber}}},
  \bibinfo{journal}{Advances in Space Research} \textbf{\bibinfo{volume}{34}},
  \bibinfo{pages}{491} (\bibinfo{year}{2004}).

\bibitem[{\citenamefont{Ruhl et~al.}(2004)\citenamefont{Ruhl, Ade, Carlstrom,
  Cho, Crawford, Dobbs, Greer, Halverson, Holzapfel, Lantin et~al.}}]{SPT}
\bibinfo{author}{\bibfnamefont{J.}~\bibnamefont{Ruhl}},
  \bibinfo{author}{\bibfnamefont{P.}~\bibnamefont{Ade}},
  \bibinfo{author}{\bibfnamefont{J.}~\bibnamefont{Carlstrom}},
  \bibinfo{author}{\bibfnamefont{H.}~\bibnamefont{Cho}},
  \bibinfo{author}{\bibfnamefont{T.}~\bibnamefont{Crawford}},
  \bibinfo{author}{\bibfnamefont{M.}~\bibnamefont{Dobbs}},
  \bibinfo{author}{\bibfnamefont{C.}~\bibnamefont{Greer}},
  \bibinfo{author}{\bibfnamefont{N.}~\bibnamefont{Halverson}},
  \bibinfo{author}{\bibfnamefont{W.}~\bibnamefont{Holzapfel}},
  \bibinfo{author}{\bibfnamefont{T.}~\bibnamefont{Lantin}},
  \bibnamefont{et~al.}, \bibinfo{journal}{Proc. SPIE}
  \textbf{\bibinfo{volume}{5498}}, \bibinfo{pages}{11} (\bibinfo{year}{2004}).

\bibitem[{\citenamefont{{Kosowsky}}(2003)}]{ACT}
\bibinfo{author}{\bibfnamefont{A.}~\bibnamefont{{Kosowsky}}},
  \bibinfo{journal}{New Astronomy Review} \textbf{\bibinfo{volume}{47}},
  \bibinfo{pages}{939} (\bibinfo{year}{2003}).

\bibitem[{\citenamefont{{Zaldarriaga}}(2001)}]{zaleb}
\bibinfo{author}{\bibfnamefont{M.}~\bibnamefont{{Zaldarriaga}}},
  \bibinfo{journal}{\prd} \textbf{\bibinfo{volume}{64}},
  \bibinfo{pages}{103001} (\bibinfo{year}{2001}).

\bibitem[{\citenamefont{{Lewis} et~al.}(2002)\citenamefont{{Lewis},
  {Challinor}, and {Turok}}}]{LCT}
\bibinfo{author}{\bibfnamefont{A.}~\bibnamefont{{Lewis}}},
  \bibinfo{author}{\bibfnamefont{A.}~\bibnamefont{{Challinor}}},
  \bibnamefont{and} \bibinfo{author}{\bibfnamefont{N.}~\bibnamefont{{Turok}}},
  \bibinfo{journal}{\prd} \textbf{\bibinfo{volume}{65}},
  \bibinfo{pages}{023505} (\bibinfo{year}{2002}).

\bibitem[{\citenamefont{{Bunn}}(2002{\natexlab{a}})}]{bunn}
\bibinfo{author}{\bibfnamefont{E.~F.} \bibnamefont{{Bunn}}},
  \bibinfo{journal}{\prd} \textbf{\bibinfo{volume}{65}},
  \bibinfo{pages}{043003} (\bibinfo{year}{2002}{\natexlab{a}}).

\bibitem[{\citenamefont{{Bunn}}(2002{\natexlab{b}})}]{bunnerratum}
\bibinfo{author}{\bibfnamefont{E.~F.} \bibnamefont{{Bunn}}},
  \bibinfo{journal}{\prd} \textbf{\bibinfo{volume}{66}},
  \bibinfo{pages}{069902} (\bibinfo{year}{2002}{\natexlab{b}}).

\bibitem[{\citenamefont{{Bunn} et~al.}(2003)\citenamefont{{Bunn},
  {Zaldarriaga}, {Tegmark}, and {de Oliveira-Costa}}}]{bunnetal}
\bibinfo{author}{\bibfnamefont{E.~F.} \bibnamefont{{Bunn}}},
  \bibinfo{author}{\bibfnamefont{M.}~\bibnamefont{{Zaldarriaga}}},
  \bibinfo{author}{\bibfnamefont{M.}~\bibnamefont{{Tegmark}}},
  \bibnamefont{and} \bibinfo{author}{\bibfnamefont{A.}~\bibnamefont{{de
  Oliveira-Costa}}}, \bibinfo{journal}{\prd} \textbf{\bibinfo{volume}{67}},
  \bibinfo{pages}{023501} (\bibinfo{year}{2003}).

\bibitem[{\citenamefont{{Zare}}(1988)}]{zare}
\bibinfo{author}{\bibfnamefont{R.~N.} \bibnamefont{{Zare}}},
  \emph{\bibinfo{title}{{Angular momentum: Understanding spatial aspects in
  chemistry and physics}}} (\bibinfo{publisher}{John Wiley \& Sons},
  \bibinfo{year}{1988}).

\end{thebibliography}

\end{document}